\documentclass{article}
\usepackage{amsmath,cite}
\usepackage{graphicx}
\usepackage{amssymb,cite}
\usepackage{esint}
\def\bs{\boldsymbol}
\def\G{\mathcal G}
\def\C{\mathcal C}
\def\L{\mathcal L}
\makeatletter
\providecommand{\LyX}{L\kern-.1667em\lower.25em\hbox{Y}\kern-.125emX\@}
\numberwithin{equation}{section}

\makeatother

\begin{document}

\title{Spatial contraction of the Poincar\'e group and Maxwell's 
equations in the electric limit}

\author{H.T.~Reich and S.~Wickramasekara\\
Department of Physics\\
Grinnell College\\
Grinnell, IA 50112}

\maketitle
\begin{abstract}
The contraction of the Poincar\'e group with respect to the space translations subgroup gives rise to a group that bears a certain duality relation to the Galilei group, that is, the contraction limit of the Poincar\'e group with respect to the time translations subgroup. In view of this duality, we call the former the dual Galilei group. A rather remarkable feature of the dual Galilei group is that the time translations constitute a central subgroup. Therewith, in unitary irreducible representations (UIR) of the group, the Hamiltonian appears as a Casimir operator proportional to the identity $H=EI$, with $E$ (and a spin value $s$) uniquely characterizing the representation. Hence, a physical system characterized by a UIR of the dual Galilei group displays no non-trivial time evolution. Moreover, the combined $U(1)$ gauge group and the dual Galilei group underlie a non-relativistic limit of Maxwell's equations known as the electric limit. The analysis presented here shows that only electrostatics is possible for the electric limit, wholly in harmony with the trivial nature of time evolution governed by the dual Galilei group. 
\end{abstract}
\pagebreak
\section{Introduction}\label{sec1}

This paper is a sequel to a recent study on the Galilean transformation properties of 
a $U(1)$ gauge field coupled to a matter field~\cite{victorsujeev08}. In particular, 
it is shown in \cite{victorsujeev08} that local $U(1)$-gauge invariance of the 
Lagrangian density requires that the 
matter field for a particle be coupled to a gauge field and that the equations
of motion for the gauge field are the familiar Maxwell's equations whether 
the matter field describes a relativistic particle or a non-relativistic particle. 
The analysis 
of \cite{victorsujeev08} hinges on the following 
two points: 
\begin{enumerate}
\item[{(1.A)}]The wave functions $\psi(\bs{x}, t)$ inhabit a Hilbert space 
that furnishes a unitary, 
irreducible, projective, representation of the Galilei group. Such representations are 
characterized by three numbers $(m,s,w)$ which have interpretation as mass, spin and internal energy, 
respectively.  The discussion of \cite{victorsujeev08} considered the $s=0$ case. 
\item [{(1.B)}] Under the action of the Galilei group, 
the four-component gauge field $\left(A_0,\bs{A}\right)$ transforms 
as 
\begin{eqnarray}
A_0\left(\bs{x},t\right)&\to& A'_0(\bs{x},t)=A_0(\bs{x}',t')-\bs{v}\cdot R\bs{A}\left(\bs{x}',t'\right)\nonumber\\
\bs{A}\left(\bs{x},t\right)&\to& \bs{A}'\left(\bs{x},t\right)=R\bs{A}\left(\bs{x}',t'\right)\label{1.1}
\end{eqnarray}
where $t'=t-b$ and $\bs{x}'=R^{-1}\bs{x}-R^{-1}\bs{v}t-R^{-1}\left(\bs{a}-b\bs{v}\right)$. 
\end{enumerate}
It follows from \eqref{1.1} that the electric and magnetic fields, defined as usual in terms of potentials 
$\left(A_0,\bs{A}\right)$, transform under the Galilei group as 
\begin{eqnarray}
\bs{E}\left(\bs{x},t\right)&\to& \bs{E}'\left(\bs{x},t\right)=R\bs{E}\left(\bs{x}',t'\right)-\bs{v}\times R\bs{B}\left(\bs{x}',t'\right)\nonumber\\
\bs{B}\left(\bs{x},t\right)&\to&\bs{B}'\left(\bs{x},t\right)=R\bs{B}\left(\bs{x}',t'\right)\label{1.2}
\end{eqnarray}
The Galilean transformation formulas \eqref{1.2} are sometimes called the ``magnetic limit" of the Lorentz transformations since 
they  can be obtained as the $\frac{v}{c}\to0$ limit 
of the Lorentz transformation formula for the electromagnetic field tensor under the constraint 
$c\left|\bs{B}\right|>>\left|\bs{E}\right|$ \cite{lebellac}. Similarly, there is an ``electric limit'', the $\frac{v}{c}\to0$ limit 
of the Lorentz transformation formula for the electromagnetic tensor when $\left|\bs{E}\right|>>c\left|\bs{B}\right|$:
\begin{eqnarray}
\bs{E}\left(\bs{x},t\right)&\to&\bs{E}'\left(\bs{x},t\right)=R\bs{E}\left(\bs{x}',t'\right)\nonumber\\
\bs{B}\left(\bs{x},t\right)&\to&\bs{B}'\left(\bs{x},t\right)=R\bs{B}\left(\bs{x}',t'\right)+
\frac{\bs{v}}{c^2}\times R\bs{E}\left(\bs{x},t\right)\label{1.3}
\end{eqnarray}
These equations correspond to the following Galilean transformation formulas for the gauge field:
\begin{eqnarray}
A_0\left(\bs{x},t\right)&\to& A'_0(\bs{x},t)=A_0(\bs{x}',t')\nonumber\\
\bs{A}\left(\bs{x},t\right)&\to& \bs{A}'\left(\bs{x},t\right)=R\bs{A}\left(\bs{x}',t'\right)+\frac{\bs{v}}{c^2}A_0\left(\bs{x}',t'\right)\label{1.4}
\end{eqnarray}

An important conclusion of \cite{victorsujeev08} is that \emph{(1.A) implies (1.B)}. That is, 
if (1.A) holds for the matter field, then the theory has meaningful transformation properties 
under the Galilei group only for the magnetic limit of the gauge field. Therefore, 
if the equations of motions are to be Galilean invariant in the electric limit, then the condition (1.A) must not hold, 
i.e., the matter field must not transform under a unitary, irreducible, projective representation of the Galilei group. 
Instead, it was suggested in \cite{victorsujeev08} that the Hilbert space of wave functions must furnish a 
unitary, irreducible, (possibly) projective representation (UIR) of what was called the  ``dual Galilei group". 
The main technical results we report in this paper are an analysis of this ``dual Galilei group", including 
how the group results from an In\"on\"u-Wigner contraction of the Poincar\'e group with 
respect to the spatial-translation subgroup and the construction of its UIR's. 
These representations can then be combined with local $U(1)$ gauge transformations to obtain the 
electric limit equations. However, the situation here is quite different from  
the magnetic limit equations that arise from the unitary projective representations of the Galilei group:  
due to the nature of time evolution governed by the dual Galilei group, a $U(1)$ gauge theory constrained 
by UIR's of the dual Galilei group gives rise to only electrostatics. 

There have been several studies on whether or not the structure of Maxwell's equations depends on the symmetry structure 
of spacetime. Some examples are  \cite{dyson,dombey,brehme,anderson,horowitz,kapuscik,lebellac,brown,goldin,heras}. 
For a discussion of the literature  and further references, see \cite{victorsujeev08}. 

The point of view advocated here and in \cite{victorsujeev08} is that
 the transformation properties of spacetime coordinates under a symmetry group do not 
uniquely determine how various quantities, such as fields and charge-current densities, that appear 
in a given dynamical equation transform under this group. Therefore, the invariance of a dynamical equation 
under a symmetry group is in part a matter of how various dynamical variables are \emph{defined} to transform 
under the group.  

For instance, consider an arbitrary group $G$ that has realization as a group of 
homogeneous spacetime transformation matrices $D(g)$, $g\in G$:
\begin{equation}
D(g):\  x^\mu\to x'^\mu=D_{\mu\nu}(g)x^\mu\label{1.5}
\end{equation}
where we have denoted an arbitrary element of $\mathbb{R}^4$  by $x^\mu$ and 
adopted the convention that repeated indices are to be summed over, regardless of whether they appear 
as superscripts or subscripts. For notational simplicity, we define $x^0=ct$, where $c$ is a scaling constant with units of 
speed. (Despite the overt similarity of notation, we 
do not mean that ${x}^\mu$ in \eqref{1.5} is a contravariant vector with respect to the Lorentz group. We also do not mean that $c$ is 
the speed of light or another  
physical velocity invariant under the group $G$, albeit it is convenient to set $c$ to have the same numerical value as the speed of light.) 
The transformation formula \eqref{1.5} implies, by way of chain rule, that the differential operators $\frac{d}{dx^\mu}$ 
transform under $G$ as 
\begin{equation}
\frac{d}{dx'^\mu}=D^T_{\mu\nu}\left(g^{-1}\right)\frac{d}{dx^\nu}\label{1.6}
\end{equation}
where $D^T(g)$ is the transpose of the matrix $D(g)$. Now, 
recall that for a given representation $U: G\times\Phi\to\Phi$ of $G$  in a vector space $\Phi$, we can always obtain 
another representation $V: G\times\Phi^\times\to\Phi^\times$ in the dual space $\Phi^\times$ by the duality formula
\begin{equation}
\langle U(g)\phi|F\rangle=\langle\phi|V\left(g^{-1}\right)F\rangle,\quad \phi\in\Phi,\ F\in\Phi^\times\label{2.2.5}
\end{equation}
or, in more compact operator notation
\begin{equation}
V(g)=\left(U\left(g^{-1}\right)\right)^\times\label{2.2.6}
\end{equation}
where $\left(U(g)\right)^\times$ is the dual operator to $U(g)$. When $\Phi$ is a complex 
Hilbert space, the dual space $\Phi^\times$ can be identified with $\Phi$ and 
the dual operator becomes simply the adjoint, $\left(U(g)\right)^\times=\left(U(g)\right)^\dagger$.  
The property crucial to our analysis is the invariance of the bilinear form $\langle\psi|F\rangle$ under 
the two representations $U$ and $V$. It follows from \eqref{2.2.5}
\begin{equation}
\langle U(g)\phi|V(g)F\rangle=\langle\phi|F\rangle,\quad \phi\in\Phi,\ F\in\Phi^\times,\ g\in G\label{1.7}
\end{equation}

Returning to the transformations of $\mathbb{R}^4$, a 
comparison of \eqref{2.2.6} with \eqref{1.6} shows that spacetime vectors and differential operators
always transform under a group according to two different representations that are dual to each other (with 
respect to a some bilinear or sesquilinear form (e.g., inner product) as in \eqref{2.2.5}).  Let is denote the 
representation dual to \eqref{1.5} by $C$, i.e, 
\begin{equation}
C(g):=D^T\left(g^{-1}\right),\ \ g\in G\label{1.8}
\end{equation}
Since the dual 
space of $\mathbb{R}^4$  can be identified with itself, we  consider two copies $\mathbb{R}^4$, one of vectors 
transforming under the representation $D$ and the other of vectors transforming under the dual representation 
$C$. We will use superscript indices for vectors transforming under $D$ and subscript indices for those transforming 
under $C$. Therefore, we define $d_\mu:=\frac{d}{dx^\mu}$ and rewrite \eqref{1.6} simply as 
\begin{equation}
d'_\mu=C_{\mu\nu}(g)d_\nu\label{1.9}
\end{equation}
Thus, we have either 
\begin{equation}
x'^\mu=D_{\mu\nu}(g)x^\nu\ \Rightarrow\  d'_\mu=C(g)_{\mu\nu}d_\nu\label{1.10}
\end{equation}
or, with $d^\mu=\frac{d}{dx_\mu}$, 
\begin{equation}
x'_\mu=C_{\mu\nu}(g) x_\nu\ \Rightarrow\ d'^\mu=D_{\mu\nu}(g)d^\nu\label{1.11}
\end{equation}
Furthermore, according to \eqref{1.7}, we have
\begin{equation}
x'^\mu y'_\mu=x^\mu y_\mu. \label{1.12}
\end{equation} 

If $G$ is a compact Lie group, then it is possible to define a positive definite inner product with the property 
\eqref{1.12} such that $D(g)=C(g)$, i.e., $D(g)$ (and thus also $C(g)$) are orthogonal matrices. For non-compact 
groups like the homogeneous Galilei and Lorentz groups, this is not possible. However, for the Lorentz group, 
the metric $\eta_{\mu\nu}=(1,-1,-1,-1)$ provides the link between the two representations
\begin{equation}
C(g)=\eta D(g)\eta,\quad g\in SO(1,3)\label{1.12b}
\end{equation}

Even if we have decided what the group and its particular representation relevant for spacetime transformations are, 
these choices do not automatically determine the transformation properties of dynamical variables. For instance, a theory 
such as electrodynamics may contain various fields  and simply knowing how their arguments transform does 
not allow us to determine how their components transform. Thus, for instance, we may take a four component gauge field 
to furnish a $D$-representation,
\begin{equation}
A'^\mu(x)=D_{\mu\nu}(g)A^\nu(x')\label{1.13}
\end{equation}
or a $C$-representation
\begin{equation}
A'_\mu(x)=C_{\mu\nu}(g)A_\nu(x')\label{1.14}
\end{equation}
under a given group $G$. 
In each equation \eqref{1.13} or \eqref{1.14}, the argument $x'$ on the right-hand-side must be defined 
either as $x'=D\left(g^{-1}\right)x$ or as $x'=C\left(g^{-1}\right)$. The inverse group elements in the arguments are necessary 
to satisfy the homomorphism property of the representation. 

Once a choice is made between \eqref{1.13} and \eqref{1.14} so that $A$ is a vector field under $G$, 
then we can use \eqref{1.10} or \eqref{1.11} 
to build tensor fields. Thus,
\begin{equation}
F^{\mu\nu}(x)=d^\mu A^\nu(x)-d^\nu A^\mu(x)\label{1.15}
\end{equation}
is a second rank tensor under $D$-representation, whereas 
\begin{equation}
F_{\mu\nu}=d_\mu A_\nu(x) -d_\nu A_\mu(x)\label{1.16}
\end{equation}
is a second rank tensor under the $C$-representation. Clearly, we can consider higher order as well as mixed tensors. 
Then, in view of \eqref{1.12},  an equation such as 
\begin{equation}
d^\mu F_{\mu\nu}(x)=j_\nu(x)\label{1.17}
\end{equation}
can be made an invariant tensor identity with respect to the transformation group $G$.  In the same vein, 
recall that when we formulate Maxwell's equations in a gravitational field, we simply \emph{define} $F_{\mu\nu}$ to be a 
tensor under general coordinate transformations and appeal to the principle of general covariance. 

In relativistic theories, the distinction between contravariant and covariant vectors is precisely in their transformation formulas with 
respect to the dual representations of the Lorentz group, and this duality is clearly central to covariant  tensor equations. In the same spirit, an attempt to formulate a Galilean invariant dynamical theory must allow for 
both $D$ and $C$ representations of the homogeneous Galilei group.  However, while it is possible, as 
a result of \eqref{1.12b}, to cast a relativistic equation as a tensor identity with respect to either representation, 
this is not always the case for a non-relativistic equation. In the analysis presented here, it will be seen that 
both representations are necessary to formulate Galilean transformations of Maxwell's equations 
in the electric limit.

The above reasoning suggests that it may not be possible to completely determine the 
symmetry transformation group of spacetime by appealing to the invariance properties of a 
dynamical differential equation. More importantly still, it appears that the symmetry 
structure of a dynamical equation is rather devoid of physical content. However, two important constraints that 
limit the choice of possible  groups do accord physical content to the symmetry structure of spacetime. 
First, a physical system may be described 
by more than one tensor equation--for instance, in addition to \eqref{1.17}, we also have 
\begin{equation}
\epsilon_{\mu\nu\rho\sigma}d^\nu F^{\rho\sigma}=0\label{1.18}
\end{equation}
in electrodynamics. The simultaneous requirement that both \eqref{1.17} and \eqref{1.18} be tensor 
identities under (possibly different representations of) the same group can limit the range of possibilities for $G$. 
Second, if the dynamical variables such as $F^{\mu\nu}$ and $j^\mu$ are obtained as 
functions of quantum mechanical state vectors (wave functions), then the transformation 
properties of these dynamical variables will be in part determined by the 
unitary representations, which are often infinite dimensional, of the group. 
These representations may limit the freedom to define various dynamical variables as tensors under a given group, 
thereby rendering some spacetime symmetry groups as 
physically untenable. A case in point is again Maxwell's equations. In the analysis 
of \cite{victorsujeev08}, which relies on the $U(1)$ gauge invariance to introduce the fields $A_\mu$, 
the charge and current densities were defined in terms of a matter field that transforms under a UIR 
the Galilei group. A charge-current density field obtained from a representation of the Galilei group 
clearly cannot be defined a four vector under, for instance,  the homogeneous 
Lorentz group. A similar argument applies to the Feynman-Dyson study \cite{dyson}. 
In this paper, dealing with the electric limit of Maxwell's equations, we take the matter field to furnish a unitary irreducible 
representation of the dual Galilei group, defined below. A matter field transforming under such a representation 
does not define a non-trivial current density--and therewith, the resulting theory is not a dynamical theory. 
The conclusion to be drawn is that the limits imposed by the representation theory of quantum mechanics
may make it impossible to maintain both the physical content of an equation and its covariance under a given group. 

The organization of this paper is as follows: In the remainder of this Introduction, we will define what we mean 
by $C$ and $D$ representations of the homogeneous Galilei transformations and introduce the dual Galilei group. 
 In Section~\ref{sec2}, we will give an expanded summary of the results 
of~\cite{victorsujeev08} as a way of providing a context for this paper. In Section~\ref{sec3}, we examine the In\"on\"u-Wigner contractions of the Poincar\'e group 
leading to the ``dual Galilei group" and construct its UIR's. We will then use 
these representations  and gauge 
transformations  to study Maxwell's equations in the electric limit in Section \ref{sec4}.  We will make a few concluding remarks in Section~\ref{sec5}.

 \subsection{Galilei group: Definitions and preliminaries}\label{sec2.2.1}
The Galilei group $\G(1:3):=\left\{(b,\bs{a},\bs{v},R)\right\}$, where 
$R$ is an orthogonal matrix, $\bs{a}$ and $\bs{v}$ are vectors under $R$ and $b$ is a scalar, 
 is the group defined by the product rule 
\begin{equation}
\left(b_{2},\mathbf{a}_{2},\mathbf{v}_{2},R_{2}\right)\left(b_{1},\mathbf{a}_{1},\mathbf{v}_{1},R_{1}\right)=
\left(b_{2}+b_{1},\bs{a}_2+R_{2}\mathbf{a}_{1}+b_{1}\mathbf{v}_{2},\bs{v}_2+R_{2}\mathbf{v}_{1},R_{2}R_{1}\right)\label{2.2.1}
\end{equation}
There exists a natural realization of the Galilei group as a transformation group  on the spacetime $\mathbb{R}^4$:
\begin{equation}
\left(\begin{array}{c}
\bs{x}\\
ct\\
1
\end{array}\right)
\to
\left(\begin{array}{c}
\bs{x}'\\
ct'\\
1
\end{array}\right)
=\left[\begin{array}{ccc}
R & \bs{\beta} & \mathbf{a}\\
0 & 1 & cb\\
0 & 0 & 1\end{array}\right]
\left(\begin{array}{c}
\bs{x}\\
ct\\
1
\end{array}\right)\label{2.2.2}
\end{equation}
where $\bs{\beta}=\frac{\bs{v}}{c}$. (Again, recall that $c$ is only a scaling constant here and we do not require, for instance, $\beta<1$ as in special relativity.) 
It follows from \eqref{2.2.2} that $\bs{a}$ and $b$ are space and time translations, respectively, $R$ is a rotation and 
$\bs{v}$ (or $\bs{\beta}$) is a boost transformation. Setting $\bs{a}=0$ and $b=0$, we obtain the homogeneous Galilei transformations 
\begin{equation}
\left(\begin{array}{c}
\bs{x}'\\
ct'
\end{array}\right)=
D(g)
\left(\begin{array}{c}
\bs{x}\\
ct
\end{array}\right)
:=
\left[\begin{array}{cc}
R & \bs{\beta}\\
0 & 1\\
\end{array}\right]
\left(\begin{array}{c}
\bs{x}\\
ct
\end{array}\right)\label{2.2.3}
\end{equation}
We denote the group of homogeneous Galilei transformations $\{D(g)\}$ by $\G(1:3)_{\rm hom}$. It is 
isomorphic to the three dimensional Euclidean group, $E(3)$. The Galilei group \eqref{2.2.2} is the semidirect 
product of $\G(1:3)_{\rm hom}$ and the group of spacetime translations, $\G(1:3)=\left\{D(g)\right\}\rtimes T$, 
where 
\begin{equation}
T=\left\{T(g)\right\}=\left\{\left[\begin{array}{ccc}
I & \mathbf{0} & \mathbf{a}\\
0 & 1 & cb\\
0 & 0 & 1\end{array}\right]\right\}\label{2.2.4}
\end{equation}
Note that all matrices $D(g)=\left[\begin{array}{cc}
R & \bs{\beta}\\
0 & 1\\
\end{array}\right]$ leave the time-component of the $\mathbb{R}\otimes{\mathbb{R}}^3$-elements 
invariant. The notation $(1:3)$ in $\G(1:3)$ encodes this property. 

In the next section, we will define 
the dual group to $\G(1:3)_{\rm hom}$ and denote it by $\G(3:1)_{\rm hom}$ because, aside from rotations, 
this group leaves the spatial-components of $\mathbb{R}\otimes{\mathbb{R}}^3$ invariant. The group 
$\G(3:1)_{\rm hom}$ is relevant because the differential operators 
$\left(\nabla,\frac{1}{c}\frac{d}{dt}\right)$ transform under this group when spacetime vectors transform under the Galilei 
group.  In fact, it is this property that selects the magnetic limit \eqref{1.1} as the proper transformation 
rule for the electromagnetic potentials. 

\subsection{Dual Galilei group: Definitions and preliminaries}\label{sec2.2.2}
Using \eqref{2.2.6}, we define the representation of $E(3)$ dual to the one given by the matrices $D(g)$ of 
\eqref{2.2.3} as follows:
\begin{equation}
C(g):=D^{T}\left(g^{-1}\right)=
\left[\begin{array}{cc}
R & \bs{0}\\
\widehat{-R^{-1}\bs{\beta}}& 1\\
\end{array}\right]
\label{2.2.7}
\end{equation}
where  $\widehat{-R^{-1}\bs{\beta}}$ is the row vector dual to the column vector $-R^{-1}\bs{\beta}$. 
Treating the matrices $D(g)$ as acting on the spacetime manifold $\mathbb{R}\otimes{\mathbb{R}}^3$, we have 
\begin{equation}
\left(\begin{array}{c}
\bs{x}'\\
ct'
\end{array}\right)=
C(g)
\left(\begin{array}{c}
\bs{x}\\
ct
\end{array}\right)
:=
\left[\begin{array}{cc}
R & \bs{0}\\
\widehat{-R^{-1}\bs{\beta}}& 1\\
\end{array}\right]
\left(\begin{array}{c}
\bs{x}\\
ct
\end{array}\right)\label{2.2.8}
\end{equation}
Note that under the action of $C(g)$, the spatial components are simply rotated, $\bs{x}\to R\bs{x}$, while the time component 
undergoes a more complicated transformation $ct\to ct-\bs{\beta}\cdot R\bs{x}$. Hence, there is a reversal of the spatial and temporal components 
in terms of their transformation properties under the $E(3)$ representations furnished by $D(g)$ matrices and by $C(g)$ matrices. In view 
of this role reversal, we denote the group defined by \eqref{2.2.8} by
\begin{equation}
\G(3:1)_{\rm hom}:=\left\{C(g)\right\}\label{2.2.9}
\end{equation}
It is the dual representation of the homogeneous Galilei group $G(1:3)_{\rm hom}$. We denote the semidirect 
product of \eqref{2.2.9} with the spacetime translation group \eqref{2.2.4} by 
\begin{equation}
\G(3:1):=\G(3:1)_{\rm hom}\rtimes T\label{2.2.10}
\end{equation}
As a transformation group on ${\mathbb{R}}^4$,  the dual group\footnote{Note that it is only the homogeneous  groups \eqref{2.2.3} and \eqref{2.2.7} that fulfill duality relation \eqref{2.2.6}, not the inhomogeneous groups  \eqref{2.2.11} and \eqref{2.2.2}. Nevertheless, for simplicity of nomenclature, we refer to \eqref{2.2.11} as the ``dual Galilei group",  acknowledging that it is in fact the direct product of the dual of the homogenous Galilei group and the translation group of Euclidean spacetime.} \eqref{2.2.10} has realization as 
\begin{equation}
\left(\begin{array}{c}
\bs{x}\\
ct\\
1
\end{array}\right)
\to
\left(\begin{array}{c}
\bs{x}'\\
ct'\\
1
\end{array}\right)
=\left[\begin{array}{ccc}
R & {0}& \mathbf{a}\\
\widehat{-R^{-1}\bs{\beta}} & 1 & cb\\
0 & 0 & 1\end{array}\right]
\left(\begin{array}{c}
\bs{x}\\
ct\\ 
1
\end{array}\right)\label{2.2.11}
\end{equation}
It is noteworthy that when spacetime elements $(\bs{x},ct)$ transform under the Galilei group $\G(1:3)$ or the 
dual group $\G(3:1)$, the differential  operators $(\nabla,\frac{1}{c}\frac{d}{dt})$ transform under $\G(3:1)_{\rm hom}$ or 
$\G(1:3)_{\rm hom}$, respectively.
As shown in \cite{victorsujeev08} in detail and discussed briefly in the next Section, both the 
Galilei group $\G(1:3)$ and the $E(3)$ representation $\G(3:1)_{\rm hom}$ 
play a crucial role in electrodynamics in the magnetic limit.  On the other hand,
the dual group $\G(3:1)$ and the homogeneous Galilei group $\G(1:3)_{\rm hom}$ play a role 
in the Galilean invariance of Maxwell's equations in the electric limit, the discussion of Sections \ref{sec3} and 
\ref{sec4}. It is also shown in 
Section \ref{sec3} that the two groups $\G(1:3)$ and $\G(3:1)$ come about from two different contractions of the 
Poincar\'e group, the Galilei group $\G(1:3)$ from a contraction with respect to the time translation 
subgroup and the dual Galilei group $\G(3:1)$ from a contraction with respect to the 
space translation subgroup.

\section{ Maxwell's equations in the magnetic limit}\label{sec2}

\subsection{$\bs{U(1)}$-gauge covariant form of the Schr\"odinger equation}\label{sec2.1}
The symmetry properties of a physical system under local gauge transformations can be  easily 
accommodated in the Lagrangian formalism. Therefore, the analysis of \cite{victorsujeev08} begins with the 
Lagrangian that yields the Schr\"odinger equation for a free particle. Setting $\hbar=1$, a unit convention we will use 
throughout the paper, we have:
\begin{eqnarray}
L(\psi(t))&=&\frac{1}{2}\langle\psi(t)|H\psi(t)\rangle+\frac{1}{2}\langle H\psi(t)|\psi(t)\rangle-\frac{1}{2m}\langle\bs{P}\psi(t)|\bs{P}\psi(t)\rangle\nonumber\\
&=&\langle\psi(t)|H\psi(t)\rangle-\frac{1}{2m}\langle\psi(t)|\bs{P}^2\psi(t)\rangle\nonumber\\
&=&\langle\psi|H\psi\rangle-\frac{1}{2m}\langle\psi|\bs{P}^2\psi\rangle
\label{2.1}
\end{eqnarray}
The second and third equalities of \eqref{2.1} follow from the self-adjointness of operators $H$ and $\bs{P}$, 
the unitarity of the time evolution operators $U(t)$, in turn a consequence of 
the self-adjointnes of $H$, and the commutativity of $\bs{P}$ (and trivially, of $H$)  
with $U(t)$. We have defined $|\psi(t)\rangle:=U(t)|\psi\rangle$. Since the Lagrangian is time 
independent, the variations of the action $I=\int_{t_1}^{t_2} dt L$ are proportional to the
 variations of the functional $|\psi\rangle\to L$ due to small  perturbations of the state vector 
$|\psi(t)\rangle\to |\psi(t)\rangle+|\delta\psi(t)\rangle$. To first order in $|\delta\psi(t)\rangle$, we have 
\begin{eqnarray}
\delta L&=&\langle\delta\psi(t)|H\psi(t)\rangle-\frac{1}{2m}\langle\delta\psi(t)|\bs{P}^2\psi(t)\rangle
+\langle\psi(t)|H\delta\psi(t)\rangle-\frac{1}{2m}\langle\psi(t)|\bs{P}^2\delta\psi(t)\rangle\nonumber\\
&=&\langle\delta\psi(t)|H\psi(t)\rangle-\frac{1}{2m}\langle\delta\psi(t)|\bs{P}^2\psi(t)\rangle+{\rm C.C.}\label{2.2}
\end{eqnarray}
where ${\rm C.C.}$ indicates the complex conjugate of the preceding terms. 
If $L$ acquires a local extremum at $|\psi(t)\rangle$, i.e., $\delta L=0$ for arbitrary $\left|\delta\psi(t)\right.\rangle$, then we 
have 
\begin{equation}
H|\psi(t)\rangle-\frac{1}{2m}\bs{P}^2|\psi(t)\rangle=0,\label{2.3}
\end{equation}

The Lagrangian \eqref{2.1} is invariant under any unitary transformation that commutes with the Hamiltonian 
$H$ and momenta $\bs{P}$. In particular, in \eqref{2.1} we have made use of the commutativity of the unitary time evolution 
operators $U(t)$, a property that follows from the structure of the Galilei group discussed below. The Hamiltonian $H$ 
is the generator of $U(t)$, i.e., 
\begin{equation}
U(t)H=HU(t)=i\frac{dU(t)}{dt}\label{2.4}
\end{equation} 
Therefore, from \eqref{2.3} and \eqref{2.4}, we have  the equation of motion 
\begin{equation}
i\frac{d}{dt}|\psi(t)\rangle-\frac{1}{2m}\bs{P}^2|\psi(t)\rangle=0\label{2.6}
\end{equation}
the familiar Schr\"odinger's equation for a free particle (whose internal energy has been set to be zero). 
In contrast, in the case of local gauge transformations, the state vectors are transformed by unitary operators that do not 
commute with the operators $H$ and $\bs{P}$. By local gauge transformations we mean transformations of state vectors 
under  a unitary representation of the 
map group of a given compact Lie group $G$. We define the map group as the group of functions, sufficiently smooth, 
from a domain $\Omega$ into the compact group $G$.  The set of such functions becomes a group under the point-wise 
product
\begin{equation}
\left(f_2*f_1\right)(\varphi):=f_2(\varphi)*f_1(\varphi), \quad \text{for all}\ \varphi\in\Omega\label{2.7}
\end{equation}
where the $*$ on the right-hand-side is the product law of $G$. The standard strategy to obtain a Lagrangian and 
equations of motion invariant 
under local gauge transformations $U$ is to rewrite \eqref{2.1} with $H$ and $\bs{P}$ replaced by new operators $D_0$ and 
$\bs{D}$, constructed so as to commute with $U$: 
\begin{equation}
L(\psi)=\frac{1}{2}\langle\psi|D_0\psi\rangle+\frac{1}{2}\langle D_0
\psi|\psi\rangle-\frac{1}{2m}\langle\bs{D}\psi|\bs{D}\psi\rangle\label{2.8}
\end{equation}
with 
\begin{equation}
UD_0U^\dagger=D_0,\quad U\bs{D}U^\dagger=\bs{D}\label{2.9}
\end{equation}
In place of the free-particle equation  \eqref{2.3}, we then have 
\begin{equation}
D_0|\psi\rangle-\frac{1}{2m}\bs{D}^2|\psi\rangle=0\label{2.10}
\end{equation}
Since the defining criterion is \eqref{2.9}, the form of operators $D_0$ and $\bs{D}$ depends 
on the particular compact gauge group of interest. Because the underlying notions of locality can be 
easily implemented in a suitable position representation, it is common 
to take the domain $\Omega$ of the map group functions to be the entire spacetime manifold. 
Therefore, restricting the discussion to a spinless particle,
 the natural representation of the Hilbert space is the $L^2$-function space on $\mathbb{R}^3$, 
 the spectra of the position operators. 
 
For electrodynamics, the compact target group is $U(1)$. In the position representation, 
 the unitary operators furnishing representation of the the map group of $U(1)$ have  the form 
\begin{eqnarray}
\langle\bs{x}|U(\lambda)\psi(t)\rangle&=&U\left(\lambda(\bs{x},t)\right)\langle{\bs{x}}\left|\right.\psi(t)\rangle
\nonumber\\
&=&e^{-i\lambda(\bs{x},t)}\psi\left(\bs{x},t\right)\label{2.11}
\end{eqnarray}
From \eqref{2.11}, \eqref{2.4} and the form of the operators $\bs{P}$ in the position representation, we
obtain
\begin{eqnarray}
\langle{\bs{x}}\left|U(\lambda)HU^\dagger(\lambda)\right|\psi(t)\rangle&=&i\frac{d}{dt}\psi(\bs{x},t)+
i\left(\frac{dU(\lambda)}{dt}\right)U^\dagger(\lambda)\psi(\bs{x},t)\nonumber\\
\langle{\bs{x}}\left|U(\lambda)\bs{P}U^\dagger(\lambda)\right|\psi(t)\rangle&=&-i\nabla\psi(\bs{x},t)-i\left(\nabla U(\lambda)\right)U^\dagger(\lambda)\psi(\bs{x},t)\label{2.12}
\end{eqnarray}
where, for the sake of notational simplicity, we have suppressed the dependence of the  $\lambda$ on $\left(\bs{x},t\right)$ 
on the right-hand-side.  We see that the generators of time and space translations of the Galilean algebra do not commute with the 
map group action. Therefore, in order to ensure invariance of the Lagrangian under local $U(1)$-gauge transformations, we must look 
for operators $D_0$ and $\bs{D}$ that fulfill \eqref{2.9}. Since the trouble comes from the inhomogeneous terms of \eqref{2.12}, 
we can construct operators $D_0$ and $\bs{D}$ by simply adding to $H$ and $\bs{P}$ new operators $A_0$ and $\bs{A}$, 
\begin{eqnarray}
D_0&=&H+gA_0\nonumber\\
\bs{D}&=&\bs{P}+g\bs{A}\label{2.13}
\end{eqnarray}
where $g$ is a coupling constant, and demanding that, in the position representation, 
 $A_0$ and $\bs{A}$ act as multiplication by functions of $\bs{x}$ and $t$ 
that transform under map group operators  as 
\begin{eqnarray}
\langle\bs{x}\left|U(\lambda)A_0U^\dagger(\lambda)\right|\psi(t)\rangle&=&A_0(\bs{x},t)\psi(\bs{x},t)+
\frac{i}{g}\left(\frac{dU(\lambda)}{dt}\right)U(\lambda)^\dagger\psi(\bs{x},t)\nonumber\\
\langle\bs{x}\left|U(\lambda)\bs{A}U^\dagger(\lambda)\right|\psi(t)\rangle&=&\bs{A}(\bs{x},t)\psi(\bs{x},t)+
\frac{i}{g}\left(\nabla{U}(\lambda)\right)U(\lambda)^\dagger\psi(\bs{x},t)\nonumber\\
\label{2.14}
\end{eqnarray}

Our analysis of local gauge invariance is slightly different from the more standard treatments in that we approach 
the gauge transformation as a symmetry transformation of quantum \emph{states} whereas in standard treatments on 
relativistic quantum theories (see, for instance, \cite{ryder}) 
 gauge transformations are  considered transformations of quantum \emph{fields}. 

We can substitute operators \eqref{2.13} in \eqref{2.8} and \eqref{2.10} to obtain the Lagrangian and 
 equation of motion  that are invariant under local $U(1)$-gauge transformations. In particular, 
in the position representation, we have the Lagrangian
\begin{eqnarray}
L&=&\int d^3x\, \frac{i}{2}\psi^*(\bs{x},t)\left(\frac{d}{dt}+igA_0(\bs{x},t)\right)\psi\left(\bs{x},t\right)\nonumber\\
&&\ -\frac{i}{2}\psi(\bs{x},t)\left(\frac{d}{dt}-igA_0(\bs{x},t)\right)\psi^*\left(\bs{x},t\right)\nonumber\\
&&\ -\frac{1}{2m}\Bigl(\nabla+ig\bs{A}\left(\bs{x},t\right)\Bigr)\psi(\bs{x},t)\cdot\Bigl(\nabla-ig\bs{A}(\bs{x},t)\Bigr)\psi^*(\bs{x},t)
\label{2.15}
\end{eqnarray}
and the equation of motion 
\begin{equation}
i\left(\frac{d}{dt}+igA_0(\bs{x},t)\right)\psi(\bs{x},t)=-\frac{1}{2m}\Bigl(\nabla+ig\bs{A}(\bs{x},t)\Bigr)^2\psi(\bs{x},t)
\label{2.16}
\end{equation}

The transformation formulas \eqref{2.14} show that fields $\bs{E}(\bs{x},t)$ and $\bs{B}(\bs{x},t)$, defined as usual 
 by
\begin{eqnarray}
\bs{E}\left(\bs{x},t\right)&:=&\left(\nabla A_0(\bs{x},t)-\frac{d}{dt}\bs{A}(\bs{x},t)\right)\nonumber\\
\bs{B}\left(\bs{x},t\right)&:=&\nabla\times\bs{A}(\bs{x},t),\label{2.17}
\end{eqnarray}
are also invariant under local $U(1)$-gauge transformations.  Therefore, we can add an arbitrary, differentiable  
function $f(\bs{E},\bs{B})$ (that is also integrable with respect to $d^3 x$) of variables \eqref{2.17} to the integrand of \eqref{2.15} to obtain the most general $U(1)$-gauge invariant Lagrangian density: 
\begin{eqnarray}
{\cal L}&=&\frac{i}{2}\psi^*\left(\frac{d}{dt}+igA_0\right)\psi
 -\frac{i}{2}\psi\left(\frac{d}{dt}-igA_0\right)\psi^*\nonumber\\
&&\ -\frac{1}{2m}\left(\nabla+ig\bs{A}\right)\psi\cdot\left(\nabla-ig\bs{A}\right)\psi^*+f\left(\bs{E},\bs{B}\right)
\label{2.17b}
\end{eqnarray}
where we have suppressed the arguments $\bs{x}$ and $t$. 
From such a Lagrangian, by way of Euler-Lagrange 
equations, we obtain the equations of motion for the gauge field $(A_0,\bs{A})$:
\begin{eqnarray}
\nabla\cdot\nabla_{\bs{E}}f&=&-g\psi^*\psi\nonumber\\
\frac{d}{dt}\nabla_{\bs{E}}f+\nabla\times\nabla_{\bs{B}}f&=&\frac{ig}{2m}\Bigl(\psi\left(\nabla-ig\bs{A}\right)\psi^*-\psi^*\left(\nabla+ig\bs{A}\Bigr)\psi\right)
\label{2.18}
\end{eqnarray}
where 
\begin{equation}
\nabla_{\bs{E}}f:=\left(\frac{\partial{f}}{\partial{E_1}},\frac{\partial{f}}{\partial{E_2}},\frac{\partial{f}}{\partial{E_3}}\right),\quad 
\nabla_{\bs{B}}f:=\left(\frac{\partial{f}}{\partial{B_1}},\frac{\partial{f}}{\partial{B_2}},\frac{\partial{f}}{\partial{B_3}}\right)\label{2.19}
\end{equation}

Note that the two homogeneous Maxwell's equations are trivial consequences of the definitions \eqref{2.17}. The two 
inhomegenous equations can be obtained from \eqref{2.19} by letting 
\begin{equation}
f\left(\bs{E},\bs{B}\right)=\frac{g^2}{2c}\left(\bs{E}^2-c^2\bs{B}^2\right)\label{2.20}
\end{equation}
Therewith, we obtain
\begin{eqnarray}
\nabla\cdot\bs{B}&=&0\nonumber\\
\nabla\times\bs{E}+\frac{d}{dt}\bs{B}&=&0\nonumber\\
\nabla\cdot\bs{E}&=&\frac{c}{g^2}\rho\nonumber\\
c^2\nabla\times\bs{B}-\frac{d}{dt}\bs{E}&=&\frac{c}{g^2}\bs{j}\label{2.21}
\end{eqnarray}
where 
\begin{eqnarray}
\rho&=&-g\psi^*\psi\nonumber\\
\bs{j}&=&-\frac{ig}{2m}\Bigl(\psi\left(\nabla-ig\bs{A}\right)\psi^*-\psi^*\left(\nabla+ig\bs{A}\Bigr)\psi\right)\label{2.22}
\end{eqnarray}

These equations are to be supplemented with the equation of motion for the matter field, \eqref{2.16}. (This is the 
equation that takes the place of $\bs{F}=m\bs{a}$, where $\bs{F}$ is the Lorentz force $\bs{F}=q\left(\bs{E}+\bs{v}\times\bs{B}\right)$ 
in the classical theory.) With the aid of \eqref{2.16} or directly from \eqref{2.21}, we can show that $\rho$ and $\bs{j}$ defined by
\eqref{2.22} fulfill the continuity equation:
\begin{equation}
\frac{d}{dt}\rho+\nabla\cdot\bs{j}=0\label{2.23}
\end{equation}

Therewith, we see that Maxwell's equations can be obtained from the requirement that the Lagrangian for the Schr\"odinger 
equation be invariant under local $U(1)$ gauge transformation. This a rather surprising result, 
 rather similar to the one obtained by Feynman and Dyson \cite{dyson}. It leads to a natural query about the 
transformations of  Maxwell's equations under the  Galilei group.  In the next subsection, we will discuss the Galilei group and its 
representations. In subsection \ref{sec2.3}, we will summarize the Galilean transformations of 
the magnetic limit Maxwell equations.

\subsection{Unitary, irreducible, projective representations of the Galilei group $\bs{\G(1:3)}$}\label{sec2.2.3}
The method of induced representations of Wigner and Mackey \cite{wigner, mackey} can be used to construct all 
unitary, irreducible, projective representations of the Galilei group \cite{leblond, IW53,bargmann}. Here, we simply give how the state 
vectors transform under the unitary operators $U(g),\ g\in\G(1:3)$, that furnish the representation defined by 
$m$ and $s$. (Without loss of generality, the internal energy $w$ can be set to zero for an irreducible representation.) 
In the momentum representation,
\begin{equation}
\left(U(g)\psi\right)\left(\bs{p},s_3\right)=e^{-i\left(\frac{1}{2}m\bs{a}\cdot\bs{v}+\bs{a}\cdot\bs{p}'-bE'\right)}
\sum_{s'_3}D^s(R)_{s_3s'_3}\psi\left(\bs{p}',s'_3\right)\label{2.2.21}
\end{equation}
where $D^s$ are the $(2s+1)\times(2s+1)$-matrices that furnish a unitary irreducible representation of the rotation group and 
\begin{equation}
\bs{p}'=R^{-1}\bs{p}-mR^{-1}\bs{v},\quad E'=E+\bs{v}\cdot\bs{p}+\frac{1}{2}m\bs{v}^2. \label{2.2.22}
\end{equation}
For a spinless particle, to which we restrict our discussion, the transformation formula reduces to
\begin{equation}
\left(U(g)\psi\right)\left(\bs{p}\right)=e^{-i\left(\frac{1}{2}m\bs{a}\cdot\bs{v}+\bs{a}\cdot\bs{p}'-bE'\right)}
\psi\left(\bs{p}'\right)\label{2.2.23}
\end{equation}
with $\bs{p}'$ and $E'$ given still by \eqref{2.2.22}. In the position representation of a spinless particle, 
the action of $U(g)$ on the wavefunctions is 
\begin{equation}
\left(U(g)\psi\right)\left(\bs{x},t\right)=e^{-im\left(\frac{1}{2}\bs{v}^2t'-\bs{v}\cdot R\bs{x}'+C\right)}
\psi\left(\bs{x}',t'\right)\label{2.2.24}
\end{equation}
where the integration constant $C=-\frac{1}{2}\bs{a}\cdot\bs{v}+\frac{1}{2}b\bs{v}^2$, and 
\begin{equation}
\left(\begin{array}{c}
\bs{x}'\\
ct'\\
1
\end{array}\right)
=\left[\begin{array}{ccc}
R^{-1} & -R^{-1}\bs{\beta} & -R^{-1}\left(\bs{a}-cb\bs{\beta}\right)\\
0 & 1 &- cb\\
0 & 0 & 1\end{array}\right]
\left(\begin{array}{c}
\bs{x}\\
ct\\
1
\end{array}\right)\label{2.2.25}
\end{equation}
In both position and momentum representations, the arguments of the wavefunction $\psi$ on the right-hand-side are defined 
in terms those on the left-hand-side by the inverse of Galilean transformations \eqref{2.2.2}. This is necessary to ensure that the 
operators defined by \eqref{2.2.22}, \eqref{2.2.23} and \eqref{2.2.24} fulfill the group homomorphism property.  When 
$(\bs{x},ct)$ transform under inverse Galilei transformations 
as in \eqref{2.2.25}, the differential operators $\left(\nabla,\frac{1}{c}\frac{d}{dt}\right)$ transform as 
\begin{equation}
\left(
\begin{array}{c}
\nabla\\
\frac{1}{c}\frac{d}{dt}
\end{array}
\right)
=\left[
\begin{array}{cc}
R&0\\
\widehat{-R^{-1}\bs{\beta}}&1
\end{array}
\right]
\left(
\begin{array}{c}
\nabla'\\
\frac{1}{c}\frac{d}{dt'}
\end{array}
\right)\label{2.2.26}
\end{equation}

The unitary operators $U(g)$ defined by any of the formulas \eqref{2.2.21}, \eqref{2.2.23} and \eqref{2.2.24} furnish a \emph{projective} 
representation of the Galilei group. Recall that a projective representation of a group is one where the operators $U(g)$ fulfill the 
composition law
 \begin{equation}
U(g_2)U(g_1)=e^{-i\omega(g_2,g_2)}U(g_2g_2)\label{2.2.27}
\end{equation}
for some real valued function $\omega:\ G\otimes G\to\mathbb{R}$ determined, though not uniquely, 
 by the structure of the group. Such non-trivial phase factors exist either because the group $G$ has a 
 non-trivial first homotopy group or because its Lie algebra admits central charges. Both the Poincar\'e group 
 and the Galilei group  have $\mathbb{Z}_2$ as their homotopy group 
 which results in $e^{-i\omega(g_2,g_1)}=\pm1$. These projective representations are equivalent 
 to true (vector) representations of the universal covering group of the group, obtained by essentially 
 replacing the rotation subgroup by $SU(2)$. While  the projective representations of the Poincar\'e group
 are entirely of this topological origin, the Galilei group also has projective representations of algebraic 
 origin due to the existence of nontrivial central charges. Transformation formulas 
 \eqref{2.2.21}, \eqref{2.2.23} and \eqref{2.2.24} define such projective representations for which 
 the function $\omega$ in the phase factor  may be chosen to be $\omega(g_2,g_1)=m\gamma(g_2,g_1)$, 
 where 
 \begin{equation}
 \gamma(g_2,g_1)=\frac{1}{2}\left(\bs{a}_2\cdot R_2\bs{v}_1-\bs{v}_2\cdot R_1\bs{a}_1
 +b_1\bs{v}_2\cdot R_2\bs{v}_1\right)\label{2.2.28}
 \end{equation}
and $m$ is the same arbitrary real number as that which appears in \eqref{2.2.21}-\eqref{2.2.24}.

 Note that whenever we have a projective representation of a group $G$ with \eqref{2.2.27}, we may expand the group 
 into another, $\tilde{G}$, by setting 
 \begin{equation}
\tilde{G}=\left\{(\alpha,g):\ \ g\in G; \alpha\in{\mathbb{R}}\right\}\label{2.2.29}
\end{equation}
and defining a product law by 
\begin{equation}
\tilde{g}_2\tilde{g}_1=\left(\alpha_2+\alpha_1+\frac{1}{\kappa}\omega(g_2,g_1), g_2g_1\right)\label{2.2.30}
\end{equation}
where $g_2g_1$ on the right hand side is the composition of $G$,   $\frac{1}{\kappa}$ is an arbitrary real constant and 
$\omega(g_2,g_1)$ is the function defined by \eqref{2.2.27}. The set $\left\{(\alpha,e)\right\}$ is a central subgroup 
of $\tilde{G}$. Therefore, $\tilde{G}$ is  called a \emph{central extension} of $G$. It is noteworthy that the projective representation 
of $G$ with \eqref{2.2.27} is equivalent to a true representation of the central extension $\tilde{G}$, furnished 
by the operators 
\begin{equation}
U\left(\tilde{g}\right):=e^{-i\kappa\alpha}U(g)\label{2.2.31}
\end{equation}
That $U\left(\tilde{g}_2\right)U\left(\tilde{g}_1\right)=U\left(\tilde{g}_2\tilde{g}_1\right)$ readily follows from \eqref{2.2.27} and 
\eqref{2.2.30}.  

 By differentiating the group operators $U(g)$ with respect to the group parameters, we obtain a representation of the 
 Lie algebra of $G$ by self-adjoint (in general, unbounded) operators. These operators fulfill the characteristic commutation 
 relations of the group. However, if the operators $U(g)$ define a projective representation, then there appears a central charge 
 in at least one of the commutation relations, i.e., there appears a term $i\kappa I$ on the right-hand-side of at least one 
 commutation relation. Furthermore, if we construct the operator Lie algebra by differentiating the associated 
 true representation \eqref{2.2.31}, then the derivative $i\left.\frac{dU(\tilde{g})}{d\alpha}\right|_{e}$ defines a new \emph{operator} 
 $K=\kappa I$. This new operator is both a generator of a one parameter subgroup and a central element of the associated 
 enveloping algebra of $\left.dU\right|_e$. 
 
 For the Galilei group, a comparison of \eqref{2.2.28} and \eqref{2.2.30} shows that the central charge is mass $m$. The equivalent 
 true representation of the centrally extended group gives rise to a mass operator $M$ which has the form $M=mI$ in an irreducible 
 representation. A basis for the operator Lie algebra of this representation of the centrally extended group can be chosen to 
 consist of $\left\{H,\ \bs{P},\ \bs{K},\ \bs{J}, M\right\}$, the generators of time translations, space translations, boosts and rotations, 
 and the mass. They fulfill the commutation relations
 \begin{eqnarray}
\left[J_i,J_j\right]=i\epsilon_{ijk}J_k & \left[J_i,P_j\right]=i\epsilon_{ijk}P_k & \left[J_i,K_j\right]=i\epsilon_{ijk}K_k\nonumber\\
\left[P_i,P_j\right]=0 & \left[K_i,K_j\right]=0 & \boxed{\left[K_i,P_j\right]=i\delta_{ij}M}\nonumber\\
\left[J_i,H\right]=0 & \left[P_i,H\right]=0 & \left[K_i,H\right]=iP_i\nonumber\\
\left[K_i,M\right]=0 &\left[J_i,M\right]=0& \left[P_i,M\right]=\left[H,M\right]=0
\label{2.2.32}
\end{eqnarray}
For a true representation of the Galilei group, the last set of equations does not exist and the 
boxed equation of the second line becomes $\left[K_i,P_j\right]=0$. 

\subsection{Galilean transformations of the gauge field}\label{sec2.3}

 On the other hand, the transformation formula for the gauge field $(A_0, \bs{A})$ is not directly determined 
by the unitary representation associated with the matter field. The simplest choice is to take $(A_0,\bs{A})$ 
to be a vector field. However, unlike in the relativistic case, owing to the existence of two distinct representations 
$\G(1:3)_{\rm hom}$ and $\G(3:1)_{\rm hom}$ of the homogeneous Galilei transformations, we have two choices: 
\begin{enumerate}
\item The gauge field $(A_0,\bs{A})$ is a vector field under $\G(1:3)_{\rm hom}$, with the transformation formula given 
by \eqref{1.4}. This corresponds to the electric limit transformation formulas \eqref{1.3} for the electric and magnetic fields. 
\item The gauge field $(A_0,\bs{A})$ is a vector field under $\G(3:1)_{\rm hom}$, with the transformation formula given 
by \eqref{1.1}. This corresponds to the magnetic limit transformation formulas \eqref{1.2} for the electric and magnetic fields. 
\end{enumerate}
Once a choice for the transformation rule for $(A_0,\bs{A})$ is made and the transformation formula for the matter field is at hand,
the Galilean transformation properties for the charge and current densities directly follow from the definitions \eqref{2.22}. 

The free particle Lagrangian density is Galilean invariant when the matter field transforms under the unitary representation \eqref{2.2.24} and, consequently, the  differential operators $\left(\frac{d}{dt},\nabla\right)$ transform as defined by \eqref{2.2.26}. 
Therefore, since the  gauge field $(A_0, \bs{A})$ appears  as coupled to 
$\left(\frac{d}{dt},\nabla\right)$, we expect that $\left(A_0,\bs{A}\right)$ must transform the same way as $\left(\frac{d}{dt},\nabla\right)$ if the gauge invariant Lagrangian density \eqref{2.17b} or the equation
of motion \eqref{2.16} is to be Galilean invariant.  In other words, 
the natural requirement that the matter field $\psi$ transforms under the irreducible unitary representation \eqref{2.2.24}, 
in turn a consequence of the fundamental free particle Schr\"odinger equation \eqref{2.6}, selects the magnetic 
limit transformation formulas \eqref{1.1} and \eqref{1.2} for the electromagnetic potentials and fields. 

The above reasoning also leads to the  inference that, if we were to start with the electric limit transformation formulas \eqref{1.3} 
and \eqref{1.4} for the electromagnetic fields and potentials, then the matter field $\psi$ must \emph{not} transform under a unitary irreducible representation of the Galilei group $\G(1:3)$. On the contrary, since the structure of the gauge invariant 
Lagrangian density suggests that both $\left(A_0,\bs{A}\right)$ and $\left(\frac{d}{dt},\nabla\right)$ must transform 
the same way under spacetime transformations, it is natural to demand that $\left(\frac{d}{dt},\nabla\right)$ be a vector 
under $\G(1:3)_{\rm hom}$ (see \eqref{3.21b} below).  
This naturally dictates that the spacetime coordinates must transform under the dual Galilei group 
$\G(3:1)$ and therewith the matter field under a unitary, irreducible, projective representation of $\G(3:1)$. 
Constructing these representations is our main task for the next Section. Then in Section \ref{sec4} we 
use these representations to analyze the Galilean invariance of Maxwell's equations in the electric limit.

\section{Unitary, irreducible, projective representations of the dual Galilei group $\bs{\G(3:1)}$}\label{sec3}
The dual Galilei group $\G(3:1)=\left\{g\left(b,\bs{a},\bs{v},R\right)\right\}$ is the group defined 
by the product law
\begin{equation}
g\left(b_2,\bs{a}_2,\bs{v}_2,R_2\right)\cdot 
g\left(b_1,\bs{a}_1,\bs{v}_1,R_1\right)
=g\left(b_2+b_1-\frac{\bs{v}_2}{c^2}\cdot R_2\bs{a}_1,
\bs{a}_2+R_2\bs{a}_1,\bs{v}_2+R_2\bs{v}_1,R_2R_1\right)\label{3.1}
\end{equation}
It follows directly that the identity of $\G(3:1)$ is $e=(0,\bs{0},\bs{0},I)$ and the inverse of $g(b,\bs{a}\,\bs{v},R)$ is 
\begin{equation}
g^{-1}(b,\bs{a},\bs{v},R)=\left(-b-\frac{\bs{v}}{c^2}\cdot\bs{a},-R^{-1}\bs{a},-R^{-1}\bs{v},R^{-1}\right)\label{3.2}
\end{equation}
As noted above, $\G(3:1)$ has realization as  
a group of transformations on the spacetime $\mathbb{R}^4$  defined 
by  \eqref{2.2.11}. Again, a comparison between the first and second terms on the right-hand-side of equations 
\eqref{2.2.1} and \eqref{3.1} shows the role reversal of how space and time components transform 
under the two groups $\G(1:3)$ and $\G(3:1)$, a point further illustrated in the context of In\"on\"u-Wigner 
contractions of the next subsection.  

\subsection{Contractions of the Poincar\'e group to $\bs{\G(1:3)}$ and $\bs{\G(3:1)}$}\label{sec3.1}
The Poincar\'e group $\mathcal{P}=\left\{(\Lambda,a)\right\}$ is the group defined by the product law
\begin{equation}
\left(\Lambda_2,a_2\right)\left(\Lambda_1,a_1\right)=
\left(\Lambda_2\Lambda_1,a_2+\Lambda_2a_1\right)\label{2.2.13}
\end{equation}
Here, $\Lambda$ are matrices that fulfill the identity $\Lambda^T\eta\Lambda=\eta$ (Cf.~\eqref{1.12b}), where $\eta$ is the Minkowski 
metric tensor with $\eta_{00}=1,\ \eta_{11}=\eta_{22}=\eta_{33}=-1$, and $\det(\Lambda)=1$. As implied by \eqref{2.2.13}, the $a$ are vectors 
under $\Lambda$. The group $\mathcal{P}$ can be realized as a group of transformations on the spacetime given by 
\begin{equation}
x^\mu\to{x'}^\mu=\Lambda^\mu_{\ \nu}x^\nu+a^\mu\label{2.2.14}
\end{equation}
It is the accepted view that special relativistic physics acquires the form of Newtonian physics in the limit $\beta=\frac{v}{c}\to0$. 
In particular, it is expected that the Poincar\'e transformation  formula \eqref{2.2.14} reduces to the Galilean formula \eqref{2.2.2} 
in this limit. However, there are subtleties with this limiting procedure. To illustrate the point,  
consider one time dimension and one space dimension. Then, \eqref{2.2.14} reduces to the elementary formulae
\begin{eqnarray}
ct'&=&\gamma\left(ct+\beta x\right)+cb\nonumber\\
x'&=&\gamma\left(x+\beta ct\right)+a\label{2.2.15}
\end{eqnarray}
Recall $\gamma = (1-\beta^2)^{-1}$, so to first order in small $\beta$, $\gamma\approx1$  and equations \eqref{2.2.15} reduce to
\begin{eqnarray}
t'&=&t+\frac{x}{c}+b\nonumber\\
x'&=&x+vt+a\label{2.2.16}
\end{eqnarray}
These are \emph{not} the Galilean transformation formulas \eqref{2.2.2}. In fact, \eqref{2.2.2} do not define a group at all. 
We can recover the Galilean transformation rules from \eqref{2.2.15} if the small $\beta$ limit is taken under the 
additional constraint $|ct|>>|x|$, and in this sense, the Galilean transformations are a ``timelike'' limit of the Poincar\'e transformations 
for $\beta\to0$. On the other hand, if the limit $\beta\to0$ is taken under the opposite constraint $|ct|<<|x|$, then 
the relativistic formulas \eqref{2.2.15} reduce to the transformation formulas \eqref{2.2.11} for  $\G(3:1)$. 
In this sense, the dual Galilei group $\G(3:1)$ is a ``spacelike'' limit of the Poincar\'e transformations 
for $\beta\to0$.  

Since we are dealing with the limit of a sequence of groups, the proper procedure for carrying out the above limiting process 
is the In\"on\"u-Wigner contraction. We will not discuss the theory here but refer  to the original paper by In\"on\"u and Wigner \cite{IW53} as well as the subsequent generalization by Saletan \cite{saletan}. To  illustrate the essential idea, let us 
consider again the Poincar\'e transformation \eqref{2.2.15} in one time dimension and one space dimension. 
As a matrix equation, 
\begin{equation}
\left(\begin{array}{c}
x'\\
ct'\\
1
\end{array}
\right)
=
\left(\begin{array}{ccc}
\gamma & \gamma\beta & a\\
\gamma\beta & \gamma & cb\\
0 & 0 & 1\end{array}\right)
\left(\begin{array}{c}
x\\
ct\\
1\end{array}\right)\label{2.2.17}
\end{equation}
In order to obtain a given Galilei group element \eqref{2.2.2} as a limit of the Poincar\'e group element of \eqref{2.2.17}, we 
contract the latter with respect to the time translation subgroup. To that end, we introduce a dimensionless parameter $\alpha$ and scale 
all parameters of the group elements \eqref{2.2.17} by $\frac{1}{\alpha}$ except the time translation parameter: $b\to b$,  $a\to\frac{a}{\alpha}$, and $\beta\to\frac{\beta}{\alpha}$ (therewith also $\gamma\to\gamma(\alpha)=1/\sqrt{1-\left({\beta}/{\alpha}\right)^2}\Bigl.\Bigr)$. We then introduce the matrix $S=\left(\begin{array}{ccc}
\frac{1}{\alpha} & 0 & 0\\
0 & 1 & 0\\
0 & 0 & 1\end{array}\right)$ and define spacetime transformations 
\begin{eqnarray}
\left(\begin{array}{c}
x'\\
ct'\\
1
\end{array}
\right)
&=&
S^{-1}\left(\begin{array}{ccc}
\gamma(\alpha) & \gamma(\alpha)\frac{\beta}{\alpha} & \frac{a}{\alpha}\\
\gamma(\alpha)\frac{\beta}{\alpha} & \gamma(\alpha)& cb\\
0 & 0 & 1\end{array}\right)
S
\left(\begin{array}{c}
x\\
ct\\
1\end{array}\right)\nonumber\\
&=&
\left(\begin{array}{ccc}
\gamma(\alpha)& \gamma(\alpha){\beta} & {a}\\
\gamma(\alpha)\frac{\beta}{\alpha^2} & \gamma(\alpha) & cb\\
0 & 0 & 1\end{array}\right)
\left(\begin{array}{c}
x\\
ct\\
1\end{array}\right)\label{2.2.17.2}
\end{eqnarray}
It is straightforward to verify that for each fixed value $1\leq\alpha<\infty$, 
the set of transformation matrices \eqref{2.2.17.2} form a group, $G(\alpha)$, 
parametrized by $-\infty<a<\infty$, $-\infty<b<\infty$ and $-\alpha<\beta<\alpha$. 
From \eqref{2.2.17.2} and \eqref{2.2.17}, we note that
$G(\alpha)$ reduces to the Poincar\'e group for $\alpha=1$ (with $\left|\beta\right|<1$).
The  In\"on\"u-Wigner contraction is  the convergence of the sequence of groups $G(\alpha)$ to the Galilei 
group for $\alpha\to\infty$ (i.e., in the limit 
in which the similarity transformation matrix $S^{-1}$ becomes singular): 
\begin{eqnarray}
\lim_{\alpha\to\infty}S^{-1}\left(\begin{array}{ccc}
\gamma(\alpha)& \gamma(\alpha)\frac{\beta}{\alpha} & \frac{a}{\alpha}\\
\gamma(\alpha)\frac{\beta}{\alpha} & \gamma(\alpha)& cb\\
0 & 0 & 1\end{array}\right)S
&= & \underset{\alpha\to\infty}{\lim}\left(\begin{array}{ccc}
\gamma(\alpha) & \gamma(\alpha)\beta & a\\
\gamma(\alpha)\frac{\beta}{\alpha^2} & \gamma(\alpha)& cb\\
0 & 0 & 1\end{array}\right)\nonumber \\
 & = & \left(\begin{array}{ccc}
1 & \beta & a\\
0& 1 & cb\\
0 & 0 & 1\end{array}\right)
\label{2.2.18}
\end{eqnarray}
where now $-\infty<\beta<\infty$. These are the matrices of usual Galilei transformations in one spatial dimension. The generalization to 
three spatial dimensions resulting  in the matrices of \eqref{2.2.2} is straightforward. 

Note that the In\"on\"u-Wigner contraction described above is carried out under the requirement that time translations are not affected 
by the process, $b\to b$. On the other hand, it is also possible to carry out the contraction with respect to the space translation subgroup, i.e., under the requirement $\bs{a}\to\bs{a}$. Not surprisingly, such a spatial contraction yields the dual Galilei 
group $\G(3:1)$. Once again, to illustrate the procedure for one spatial and one time dimension, we let 
$b\to\frac{b}{\alpha}$, $\beta\to\frac{\beta}{\alpha}$, $a\to a$ and define a matrix $S=\left(\begin{array}{ccc}
1& 0 & 0\\
0 & \frac{1}{\alpha} & 0\\
0 & 0 & 1\end{array}\right)$.  The In\"on\"u-Wigner contraction is then given by
\begin{eqnarray}
\lim_{\alpha\to\infty}S^{-1}\left(\begin{array}{ccc}
\gamma(\alpha) & \gamma(\alpha)\frac{\beta}{\alpha} & a\\
\gamma(\alpha)\frac{\beta}{\alpha} & \gamma(\alpha) & c\frac{b}{\alpha}\\
0 & 0 & 1\end{array}\right)S
&=& \underset{\alpha\to\infty}{\lim}\left(\begin{array}{ccc}
\gamma(\alpha) & \gamma(\alpha)\frac{\beta}{\alpha^2} & a\\
\gamma(\alpha)\beta & \gamma(\alpha) & cb\\
0 & 0 & 1\end{array}\right)\nonumber \\
 & = & \left(\begin{array}{ccc}
1 & 0 & a\\
\beta & 1 & cb\\
0 & 0 & 1\end{array}\right)
\label{2.2.19}
\end{eqnarray}
or equivalently\begin{eqnarray}
x' & = & x+ a\nonumber \\
ct' & = & ct+\beta x+ cb\label{2.2.20}
\end{eqnarray}
By a straightforward generalization to three spatial dimensions and including rotations, we obtain the matrices \eqref{2.2.11} of 
the dual Galilei group $\G(3:1)$. The two contractions with respect to the 
time translation subgroup and space translation subgroup 
are the precise mathematical constructs  for the heuristic limits of \eqref{2.2.15} under requirements 
$|ct|>>|x|$ and $|ct|<<|x|$ leading to $\G(1:3)_{\rm hom}$ and $\G(3:1)_{\rm hom}$, respectively. 

\subsection{Lie aglebra of $\bs{\G(3:1)}$}\label{sec3.2}

The parameterization of $\G(3:1)$ suggests that a 
basis for the Lie algebra of $\G(3:1)$ may be denoted by $\chi_\alpha$,  where 
$\alpha=b,\bs{a},\bs{v},\bs{\theta}$.  A realization of the $\chi_\alpha$ may be obtained 
by directly differentiating the transformation matrices \eqref{2.2.11}. Alternatively, we 
may consider the natural representation $\left(U(g)f\right)(\bs{x},t)=
f\left(g^{-1}\left(\bs{x},t\right)\right)$, where $f$ is an element of a suitably defined 
function space on $\mathbb{R}^4$, and obtain a realization 
of the $\chi_\alpha$ by differentiating the operators $U(g)$ with respect to group 
parameters. In this case, we have the realization 
\begin{eqnarray}
\chi_b&=&-\frac{d}{dt}\nonumber\\
\chi_{a_i}&=&-\frac{d}{dx_i}\nonumber\\
\chi_{v_i}&=&\frac{x_i}{c}\frac{d}{dt}\nonumber\\
\chi_{\theta_i}&=&-\epsilon_{ijk}x_j\frac{d}{dx_k}\label{3.3}
\end{eqnarray}
The commutation relations characteristic of the Lie algebra of $\G(3:1)$ can be directly 
obtained from \eqref{3.3}:
\begin{eqnarray}
\left[\chi_{\theta_i},\chi_{\theta_j}\right]=\epsilon_{ijk}\chi_{\theta_k} &  
\left[\chi_{\theta_i},\chi_{a_j}\right]=\epsilon_{ijk}\chi_{a_k}&
\left[\chi_{\theta_i},\chi_{v_j}\right]=\epsilon_{ijk}\chi_{v_k}\nonumber\\
 \left[\chi_{a_i},\chi_{a_j}\right]=0&\left[\chi_{v_i},\chi_{v_j}\right]=0&
\boxed{\left[\chi_{v_i},\chi_{a_j}\right]=\frac{1}{c}\delta_{ij}\chi_b}\nonumber \\
\left[\chi_{\theta_i},\chi_b\right]=0 & \left[\chi_{a_i},\chi_b\right]=0  & \boxed{\left[\chi_{v_i},\chi_b\right]=0}
\label{3.4}
\end{eqnarray}
These commutation relations are to be compared with those of the Lie algebra of the 
Galilei group $\G(1:3)$: 
\begin{eqnarray}
\left[\chi_{\theta_i},\chi_{\theta_j}\right]=\epsilon_{ijk}\chi_{\theta_k} &  
\left[\chi_{\theta_i},\chi_{a_j}\right]=\epsilon_{ijk}\chi_{a_k}&
\left[\chi_{\theta_i},\chi_{v_j}\right]=\epsilon_{ijk}\chi_{v_k}\nonumber\\
 \left[\chi_{a_i},\chi_{a_j}\right]=0&\left[\chi_{v_i},\chi_{v_j}\right]=0&
\boxed{\left[\chi_{v_i},\chi_{a_j}\right]=0}\nonumber \\
\left[\chi_{\theta_i},\chi_b\right]=0 & \left[\chi_{a_i},\chi_b\right]=0  & \boxed{\left[\chi_{v_i},\chi_b\right]=\chi_{a_i}}
\label{3.5}
\end{eqnarray}
where $\chi_{\bs{\theta}}$, $\chi_{\bs{a}}$, $\chi_{\bs{v}}$ and $\chi_b$ are again 
the generators of rotations, space translations, time translations and boosts, respectively. 
The boxed equations in \eqref{3.4} and \eqref{3.5} highlight the differences between
 the two algebras. These differences all entail the manner in which the generator of time 
 translations appears in the commutation relations. 
 
 Either from the commutation relations \eqref{3.4} or from 
 the group product law \eqref{3.1},  
 we note the rather peculiar fact that  time translations are a central subgroup of $\G(3:1)$ 
 and therewith, the Hamiltonian a central element of its Lie algebra. This property is in sharp 
 contrast to the Galilean case. In fact, a comparison of \eqref{3.1} with \eqref{2.2.30} 
 or \eqref{3.4} with \eqref{2.2.32} shows that boosts, space translations and rotations constitute a group 
 and that the dual Galilei group 
 $\G(3:1)$ is a central extension of this group by the one-parameter time translation group. 
 In particular, note the similarities 
 between the mass operator $M$ in the centrally extended Galilean algebra  and the generator of  
 time translations $\chi_b$ in the $\G(3:1)$ Lie algebra. The boxed equations of the second line 
and the equations on the last line in the two sets \eqref{3.4} and \eqref{2.2.32} illuminate the fact 
the Hamiltonian of the $\G(3:1)$ algebra plays a role similar to that played by the mass operator
in the centrally extended $\G(1:3)$ algebra. 
 
 In view of the above observation, the first conclusion we draw is that, being a central extension itself, 
\emph{the group $\G(3:1)$ has no non-trivial projective representations of the algebraic kind}. In particular, 
 any central extension of $\G(3:1)$ is tantamount to redefining the Hamiltonian, $\tilde{H}=H+\kappa I$, 
 where $\kappa$ is a constant of the kind introduced in \eqref{2.2.31}. Owing to the property that 
 the rotation subgroup is not simply connected, there do exist projective representations of $\G(3:1)$ of 
 the topological kind. However, just as in the representations of the Galilei group 
 $\G(1:3)$ or the Poincar\'e group, the phase factors associated with these representations are 
 simply a sign, $e^{-i\omega(g_2,g_1)}=\pm1$. Therefore, allowing for this sign ambiguity, 
 we will construct  true representations of the group $\G(3:1)$, our main task in the next subsection. 
 
\subsection{Unitary irreducible representations of $\bs{\G(3:1)}$}\label{sec3.3}

We construct unitary irreducible representations (UIR's) of $\G(3:1)$ by the method of 
induced representations of Wigner and Mackey \cite{wigner,mackey,weinberg}. As in the construction of the UIR's of the 
Galilei group or the Poincar\'e group, we start with the UIR's of the maximal, Abelian, invariant 
subgroup $\C$ of $\G(3:1)$ and induce the representations of the factor group $\G(3:1)/\C$ from a 
suitable  ``little group'', the stability group of a standard set of spectral values of the generators 
of the representation of $\C$. 

The maximal Abelian invariant subgroup $\C$ consists of space and time translations, 
$\C=\left\{g(b,\bs{a},\bs{0},I)\right\}$.  Therefore, as in the case of Galilei 
group \cite{leblond}, we can try to obtain the representation of $\C$ in the momentum basis. 
To that end, let us suppose  that common eigenvectors $|\bs{p},E\rangle$ of $\bs{P}$ and $H$ exist 
as elements of a suitable vector space. Then, the one dimensional subspaces spanned by 
these eigenvectors are invariant under the representation operators $U(b,\bs{a},\bs{0},I)$. 
Therefore, 
\begin{equation}
U(b,\bs{a},\bs{0},I)|\bs{p},E,\zeta\rangle=e^{-ibE+i\bs{a}\cdot\bs{p}}|\bs{p},E,\zeta\rangle\label{3.6}
\end{equation}
where $\zeta$ denote any additional degrees of freedom needed to completely characterize these eigenvectors. 
Now, from the product law \eqref{3.1}, we have 
\begin{equation}
U(b,\bs{a},\bs{0},I)U(0,\bs{0},\bs{v},R)=U(0,\bs{0},\bs{v},R)U(b+\frac{\bs{a}\cdot\bs{v}}{c^2}, R^{-1}\bs{a},\bs{0},I)=U(b,\bs{a},\bs{v},R)\label{3.7}
\end{equation}
Applying this operator identity on the generalized eigenvector $|\bs{p},E,\zeta\rangle$, 
 we obtain
\begin{equation}
U(b,\bs{a},\bs{0},I)\Bigl(U(0,\bs{0},\bs{v},R)|\bs{p},E,\zeta\rangle\Bigr)=e^{-ibE+i\bs{a}\cdot\left(R\bs{p}-\frac{E\bs{v}}{c^2}\right)}
\Bigl(U(0,\bs{0},\bs{v},R)|\bs{p},E,\zeta\rangle\Bigr)\label{3.8}
\end{equation}
A comparison between \eqref{3.8} and \eqref{3.6} shows that the vector $U(0,\bs{0},\bs{v},R)|\bs{p},E,\zeta\rangle$ 
transforms under the translation subgroup  the same way as an eigenvector $|\bs{p}',E,\zeta'\rangle$,   
where $\zeta'$ is arbitrary, and 
\begin{equation}
\left(\begin{array}{c}
c\bs{p}'\\
E'
\end{array}
\right)
=\left(\begin{array}{cc}
R&-\bs{\beta}\\
0&1
\end{array}
\right)
\left(\begin{array}{c}
c\bs{p}\\
E
\end{array}
\right)
\label{3.9}
\end{equation}
 Therefore, $U(0,\bs{0},\bs{v},R)|\bs{p},E,\zeta\rangle$ must be a linear combination of $|\bs{p}',E,\zeta'\rangle$:
\begin{equation}
U(0,\bs{0},\bs{v},R)|\bs{p},E,\zeta\rangle=\sum_{\zeta'} D(\bs{v},R,\bs{p},E)_{\zeta'\zeta}|\bs{p}',E,\zeta'\rangle\label{3.10}
\end{equation}

Next, we must decide how the label $\zeta$ in $|\bs{p},E,\zeta\rangle$ is related to different values of $\bs{p}$ and $E$. 
To that end, we must determine how different values of $\bs{p}$ are related to one another by way of $\G(3:1)_{\rm hom}$. 
It follows from \eqref{3.9} that it is necessary to distinguish two cases: if $E\not=0$, then any $\bs{p}\in\mathbb{R}^3$ can be mapped to 
any other value $\bs{p}'\in\mathbb{R}^3$ by means of a $\G(3:1)_{\rm hom}$ element of the form 
$g(0,\bs{0},\frac{c(R\bs{p}-\bs{p}')}{E},R)$; if $E=0$, then $\bs{p}$ is constrained to a sphere with constant $p=|\bs{p}|$. 
Henceforth, we will only consider the $E\not=0$ case. 

From \eqref{3.9}, we note that there is a rotation free matrix $L(\bs{p})\in\G(3:1)_{\rm hom}$ that maps the zero vector $\bs{0}$ to any given $\bs{p}$:
\begin{equation}
\left(\begin{array}{c}
c\bs{p}\\
E
\end{array}
\right)
=
L(\bs{p})^T
\left(\begin{array}{c}
\bs{0}\\
E\end{array}
\right)
:=
\left(\begin{array}{cc}
I&\frac{c\bs{p}}{E}\\
0&1
\end{array}
\right)
\left(\begin{array}{c}
\bs{0}\\
E
\end{array}
\right)
\label{3.11}
\end{equation}
Now consider arbitrary values $(\bs{p},E)$ and elements $g(0,\bs{0},\bs{v},R)\in\G(3:1)_{\rm hom}$. With $\bs{p}'$  defined 
by \eqref{3.9}, the matrices $L(\bs{p}')$, $L(\bs{p})$ and $g(0,\bs{0},\bs{v},R)$ fulfill the following important identity:
\begin{eqnarray}
L^{-1}\left(\bs{p}')(0,\bs{0},\bs{v},R\right)L(\bs{p})&=&\left(\begin{array}{cc}
I&0\\
-\frac{c\bs{p}'}{E}&1
\end{array}
\right)
\left(\begin{array}{cc}
R&0\\
-R^{-1}\bs{\beta}&1
\end{array}
\right)
\left(\begin{array}{cc}
I&0\\
\frac{c\bs{p}}{E}&1
\end{array}
\right)\nonumber\\
&=&
\left(\begin{array}{cc}
R&0\\
0&1
\end{array}
\right)\label{3.12}
\end{eqnarray}
where the last equality follows from \eqref{3.9}. Hence, just as for the massive particle representations of the Galilei group, 
the little group is simply the rotation group. Both here and for the Galilei group, the little group elements are independent 
of  $\bs{p}$ or the boost parameter $\bs{v}$. In contrast, the little group for the massive particle representations 
of the Poincar\'e group, while isomorphic to the rotation group, is parametrized by variables that depend on the momentum and 
relevant Lorentz group element.

We can use \eqref{3.11} to \emph{define} the relationship between $|\bs{p},E,\zeta\rangle$ and $|\bs{0},E,\zeta\rangle$ so that 
$\zeta$ is independent of $\bs{p}$:
\begin{equation}
|\bs{p},E,\zeta\rangle:= N(p)U\left(L(\bs{p})\right)|\bs{0},E,\zeta\rangle=N(p)U\left(0,\bs{0},\frac{c\bs{p}}{E},I\right)|\bs{0},E,\zeta\rangle\label{3.13}
\end{equation}
where $N(p)$ is a normalization factor. 

The action of an arbitrary $U(0,\bs{0},\bs{v},R)$ on \eqref{3.13} gives
\begin{eqnarray}
U(0,\bs{0},\bs{v},R)|\bs{p},E,\zeta\rangle&=&N(p)U(0,\bs{0},\bs{v},R)U(L(\bs{p}))|\bs{0},E,\zeta\rangle\nonumber\\
&=&N(p)U(L(\bs{p}'))U\Bigl(L^{-1}(\bs{p}')(0,\bs{0},\bs{v},R)L(\bs{p})\Bigr)|\bs{0},E,\zeta\rangle\nonumber\\
&=&\frac{N(p)}{N(p')}\sum_{\zeta'}D(R)_{\zeta'\zeta}|\bs{p}',E,\zeta'\rangle\label{3.14}
\end{eqnarray}
where the last equality follows from \eqref{3.12} and definition \eqref{3.13}. Therewith, we see that the representations
of the subgroup $\G(3:1)_{\rm hom}$ are completely determined by \eqref{3.14} and the representations of the rotation group. 
These representations are well known \cite{edmunds}. As is standard, we denote the irreducible representation 
matrices of the rotation group by $D^s$, where $s=0,\frac{1}{2},1,\frac{3}{2},\cdots$. Hence, each unitary irreducible 
representation of $\G(3:1)$ is uniquely characterized by two invariants, $E$ and $s$. In order to accommodate 
these representation characterizing labels, we may denote the generalized eigenvectors by $|\bs{p},\zeta, [E,s]\rangle$. 

It only remains to determine the normalization factor $N(p)$. Since the transformation matrix of \eqref{3.9} has unit determinant, 
we can simply take the Lebesgue measure $d\bs{p}^3 dE$ as the measure invariant under $\G(3:1)$. Therefore, we may just set $N(p)=1$ 
and define the ``inner product'' as
\begin{equation}
\langle \bs{p}',\zeta',[E',s']|\bs{p},\zeta, [E,s]\rangle=\delta^3(\bs{p}'-\bs{p})\delta(E'-E)\delta_{s's}\delta_{\zeta'\zeta}\label{3.15}
\end{equation}
 
Putting \eqref{3.6}, \eqref{3.8}, \eqref{3.14} and \eqref{3.15} together, we then have the transformation formula that defines 
a UIR of $\G(3:1)$:
\begin{equation}
U(b,\bs{a},\bs{v},R)|\bs{p},\zeta,[E,s]\rangle=e^{-ibE+i\bs{a}\cdot\bs{p}'}\sum_{\zeta'}D^s(R)_{\zeta'\zeta}|\bs{p}',\zeta',[E,s]\rangle\label{3.16}
\end{equation}
with $\bs{p}'$ defined by \eqref{3.9}, $E\in{\mathbb{R}}/\left\{0\right\}$ and $s=0,\frac{1}{2}, 1,\cdots$. 

From the transformation formula \eqref{3.16} for the generalized eigenvectors, we can deduce the operators that furnish a UIR of 
$\G(3:1)$ in the bona fide Hilbert space of wave functions in the momentum representation. With 
$\psi(\bs{p},\zeta):=\langle\bs{p},\zeta,[E,s]|\psi\rangle$, we have
\begin{equation}
\Bigr(U(b,\bs{a},\bs{v},R)\psi\Bigr)\left(\bs{p},\zeta\right)=e^{-ibE+i\bs{a}\cdot\bs{p}'}\sum_{\zeta'}
D^s(R)_{\zeta\zeta'}\psi\left(\bs{p}',\zeta'\right)\label{3.17}
\end{equation}
where $c\bs{p}'=R^{-1}\left(c\bs{p}+E\bs{\beta}\right)$. 

\subsection{Generators and Lie algebra of the UIR's of $\bs{\G(3:1)}$}\label{sec3.4}
The generators of a UIR of $\G(3:1)$ can be found by the standard method: 
for each one-parameter subgroup $g(\alpha)$ of 
$\G(3:1)$, we find the associated generator by $i\left.\frac{dU(g(\alpha))}{d\alpha}\right|_{\alpha=0}$. 
Then, the set of operators $i\left.dU\right|_{e}$ furnishes a representation of the Lie algebra 
of $\G(3:1)$. A basis for this operator Lie algebra can be chosen to consist of the operators 
$i\left.\frac{dU(g(\alpha))}{d\alpha}\right|_{\alpha=0}$ for $\alpha=b,\bs{a},\bs{v},\bs{\theta}$,  
where the $\bs{\theta}$ parametrize the rotation group. In the momentum basis representation of 
\eqref{3.17}, we have the realization of the basis elements by
\begin{eqnarray}
H&=&EI\nonumber\\
\bs{P}&=&-\bs{p}\nonumber\\
\bs{K}&=&i\frac{E}{c} \frac{d}{d\bs{p}}\nonumber\\
\bs{J}&=&-i\bs{p}\times\frac{d}{d\bs{p}}+\bs{S}\label{3.18}
\end{eqnarray}
These operators fulfill the commutation relations \eqref{3.4} characteristic of $\G(3:1)$.  The unitarity 
of the representation \eqref{3.17} ensures that there exists a common dense domain of the Hilbert space  
in which all the operators of \eqref{3.18}, as well as their real linear span, may be defined as self-adjoint operators. 

The associative enveloping algebra spanned by \eqref{3.18} has two independent invariant operators: $H$ and $S^2=\Bigl(\bs{J}-\frac{\bs{K}\times{\bs{P}}}{H}\Bigr)^2$. The second operator is well defined since $H$ is invertible ($E\not=0$) 
for the representations of the kind \eqref{3.17}. In each UIR, these two invariant operators 
are proportional to the identity, their eigenvalues uniquely characterizing the representation: 
\begin{eqnarray}
H&=&EI,\ \ E\in{\mathbb{R}},\ E\not=0\nonumber\\
S^2&=&s(s+1)I,\ \ s=0,\frac{1}{2},1,\cdots\label{3.19}
\end{eqnarray}
If we accept the standard interpretation that the generator of time translations is the Hamiltonian with associated eigenvalues being 
the energy, then we see that a UIR of $\G(3:1)$ is characterized by a fixed value of energy and a fixed value of spin. Hence, as already pointed out above,  
the role 
of energy in the representations of $\G(3:1)$ is quite similar to the role of mass in the UIR's of the Galilei group $\G(1:3)$. 
The structure of the Lie algebra \eqref{3.18} also shows that there is a no natural way to introduce a mass operator that is distinct 
from the Hamiltonian. In particular, if we attempt to introduce a  mass parameter $m$ 
by way of  a projective representation along the lines the projective representations of the Galilei group, the result is simply 
a rescaling of the energy eigenvalue $E\to E+m$. 

If there exist physical systems with 
representation by the UIR's of $\G(3:1)$, then they have the remarkable feature that all physical states are eigenstates of the Hamiltonian 
(and of the square of the spin operator). In particular, this means that time translations of any state vector 
appear as a trivial phase factor, i.e., the system does not evolve at all. As we will see below, this property has implications 
for Maxwell's equations, namely  only electrostatics is possible in the electric limit. \\

\noindent{\bf Superselection Rules}\\
A natural question that may be raised at this point is if there is a superselection in $E$ for systems described by the UIR's of $\G(3:1)$, similar to 
the mass superselection rule attributed to the representations of the Galilei group $\G(1:3)$. The mass superselection rule has its origin in the projective representations of the Galilei group. From \eqref{2.2.27} and \eqref{2.2.28} we see that, in a projective representation, a set of transformations equal to the identity of the Galilei group, such as $(0,\bs{0},-\bs{v},I)(0,-\bs{a},\bs{0},I)(0,\bs{0},\bs{v},I)(0,\bs{a},\bs{0},I)=(0,\bs{0},\bs{0},I)$, is represented by an operator proportionate the identity, with a phase factor that depends on the mass: $U\Bigl((0,\bs{0},-\bs{v},I)(0,-\bs{a},\bs{0},I)(0,\bs{0},\bs{v},I)(0,\bs{a},\bs{0},I)\Bigr)=e^{im\bs{a}\cdot{\bs{v}}}I$. Therefore, a superposition of two states with different masses may acquire a relative phase factor when subjected to a set of transformations equal to  the identity of the Galilei group, i.e., such mass superpositions must not occur. However, the projective representations of the Galilei group can also be viewed as true representations of the centrally extended Galilei group. The crucial point is that as an element of the centrally extended group, with the product law defined by \eqref{2.2.30}, the above set of transformations is not equal to the identity $(0,0,0,0,I)$, but rather $(m\bs{v}\cdot\bs{a},0,0,0,I)$, which has representation by $e^{im\bs{v}\cdot{\bs{a}}}I$. On the other hand, a set of transformations equal to the identity of the extended group does not introduce a mass-dependent phase factor. Therefore, if we take the centrally extended Galilei group as our symmetry group, relative phases between different mass states do not appear as a result of the identity transformation. Furthermore, the \emph{only} difference between employing projective representations of the Galilei group or true representations of its central extension appears to be precisely the presence or absence of the mass superselection rule. In other words, the question of mass superselection rule cannot be decided on the basis of Galilean symmetry. 

Similar considerations apply to any symmetry group, including the dual Galilei group $\G(3:1)$.  If we consider the 
representations  of $\G(3:1)$ developed here as projective representations of the group consisting of rotations, boosts and spacial translations, then 
there is an $E$-superselection rule. As true representations of the group consisting of rotations, boosts and spacial translations centrally extended by time translations, there is no superselection rule in $E$. 

\subsection{UIR's of $\bs{\G(3:1)}$ in the position basis}\label{sec3.5}

In the discussion of local gauge transformations given in Section \ref{sec4} below, we will work in the position representation
of the wave functions. Therefore, we will need the explicit transformation formulas for the UIR's  of $\G(3:1)$ in the position basis. 
For the sake of simplicity, we will only consider the spin zero case.  Such a UIR will be uniquely characterized by a non-zero 
energy eigenvalue. 

From \eqref{3.18}, we see that the operators $\frac{c\bs{K}}{H}$ fulfill the canonical commutation relations 
with the momentum operators $\bs{P}$. Therefore, we may take $\bs{Q}=\frac{c\bs{K}}{H}=\frac{c\bs{K}}{E}$ 
as the position operators in the UIR, and therewith, $\{\bs{Q},H\}$ as a complete system of commuting operators (CSCO). 
Denoting the generalized common eigenvectors of this CSCO by $|\bs{x},E\rangle$, we obtain the position wave functions 
as $\psi(\bs{x}):=\langle\bs{x},E|\psi\rangle$. Following the common notation, we define $|\psi(t)\rangle=U(t,\bs{0},\bs{0},I)|\psi\rangle$
and $\psi(\bs{x},t)=\langle\bs{x},E|\psi(t)\rangle$. That is, the $t$ in $\psi(\bs{x},t)$, unlike $\bs{x}$, does not have meaning as the 
generalized eigenvalue of an operator that belongs to a CSCO. 

By unitarity of $U(g)$, we expect $\psi'=U(g)\psi$ to differ from $\psi$ only by a phase factor. This phase 
factor could depend on both spacetime points and the group element. Therefore, let
\begin{equation}
\psi'(\bs{x},t)=e^{-if(g,(\bs{x}',t'))}\psi(\bs{x}',t')\label{3.20}
\end{equation}
where
$\bs{x}'$ and $t'$ are defined by the inverse of \eqref{2.2.11}:
\begin{eqnarray}
\bs{x}'&=&R^{-1}\bs{x}-R^{-1}\bs{a}\nonumber\\
ct'&=&ct+\bs{\beta}\cdot\bs{x}-cb-\bs{\beta}\cdot\bs{a}\label{3.21}
\end{eqnarray}
It will be useful for the analysis of Section \ref{sec4} below to note that 
when $(\bs{x},t)$ transforms as in \eqref{3.21}, the differential operators $(\nabla,\frac{d}{dt})$ transform as 
\begin{eqnarray}
\nabla&=&R\nabla'+\frac{\bs{\beta}}{c}\frac{d}{dt'}\nonumber\\
\frac{d}{dt}&=&\frac{d}{dt'}\label{3.21b}
\end{eqnarray}

Now, applying the homomorphism $U(g_2)U(g_1)=U(g_2g_1)$ to \eqref{3.20} and simplifying a bit, we get 
the basic equation that must be satisfied by the function (one-cocycle) $f$:
\begin{equation}
f\left(g_2, g_1(\bs{x},t)\right)+f\left(g_1, (\bs{x},t)\right)=f\left((g_2g_1), (\bs{x},t)\right)\label{3.22}
\end{equation}
From this we conclude that $f=0$ for a true representation of $\G(3:1)$ (see remark below), and \eqref{3.20} becomes 
 \begin{equation}
\left(U(g)\psi\right)(\bs{x},t)=\psi(\bs{x}',t')\label{3.23}
\end{equation}
where $(\bs{x}',t')=g^{-1}(\bs{x},t)$ are again given by \eqref{3.21}. 

However, the representation \eqref{3.23} is not necessarily irreducible. Since, in the spinless case, 
 irreducible representations are characterized by an energy eigenvalue, we appeal to the Schr\"odinger equation:
 \begin{equation}
 i\frac{d}{dt}\psi(\bs{x},t)=H\psi(\bs{x},t)=E\psi(\bs{x},t)\label{3.34}
 \end{equation}
 This familiar eigenvalue equation tells us that the wavefunctions that inhabit an irreducible representation space 
 of $\G(3:1)$ must be of the form
 \begin{equation}
 \psi(\bs{x},t)=e^{-iEt}\tilde{\psi}(\bs{x})\label{3.25}
 \end{equation}
 where $\tilde{\psi}$ is an arbitrary, square integrable function of $\bs{x}$.  
With \eqref{3.21}, \eqref{3.23} and \eqref{3.25}, we finally have the formula for the operators that 
furnish a UIR of $\G(3:1)$ in the position basis: 
\begin{eqnarray}
\left(U(g)\psi\right)\left(\bs{x},t\right)&=&\psi(\bs{x}',t')\nonumber\\
&=&e^{-iEt'}{\tilde{\psi}}(\bs{x}')\nonumber\\
&=&e^{-iE(t-b-\frac{\bs{\beta}\cdot\bs{a}}{c})}e^{-iE\frac{\bs{\beta}\cdot\bs{x}}{c}}{\tilde{\psi}}(R^{-1}\bs{x}-R^{-1}\bs{a})\label{3.26}
\end{eqnarray}
We will make use of \eqref{3.26} in the analysis of the Galilean transformations of Maxwell's equations 
in the next Section. 
\\

\noindent{\bf Remark}\\
In passing, we note that in the general case, a non-trivial phase factor is possible in \eqref{3.20} if the 
representation is projective. In that case, \eqref{3.22} changes to 
\begin{equation}
f\left(g_2, g_1(\bs{x},t)\right)+f\left(g_1, (\bs{x},t)\right)=f\left((g_2g_1), (\bs{x},t)\right)+\omega(g_2,g_1)\label{3.27}
\end{equation}
The phase factor in the Galilei group representation formula \eqref{2.2.24} can be obtained precisely this way with 
the use of \eqref{2.2.28}. If the same procedure were to be followed for $\G(3:1)$, then the phase factor that results 
would once again be equivalent to a change of energy $E\to E+m$ in \eqref{3.26}. 

\section{Maxwell's equations in the electric limit}\label{sec4}
We now use the UIR's developed in Section \ref{sec3} to study Galilean transformation properties 
 of Maxwell's equations in the electric limit. As done in \cite{victorsujeev08} for the magnetic limit, 
 we will take a Lagrangian approach and make use of $U(1)$-gauge invariance to obtain electromagnetic 
 fields and their equations of motion. We will take the matter field to furnish a UIR of $\G(3:1)$ and, as in \eqref{1.4}, 
the gauge field to furnish a representation of $\G(1:3)_{\rm hom}$.  
  
The Lagrangian density leading to the Schr\"odinger equation \eqref{3.34} is 
\begin{equation}
\L=\frac{i}{2}\psi^*\frac{d}{dt}\psi-\frac{i}{2}\psi\frac{d}{dt}\psi^*-E\psi^*\psi\label{4.1}
\end{equation}
This Lagrangian density is clearly invariant under  a set of restricted local $U(1)$-gauge transformations 
$U(\lambda):\ \psi\to e^{-i\lambda}\psi$ for any 
$\lambda$ that is solely a function of the spatial coordinates $\bs{x}$, i.e., such local gauge transformations behave exactly as global 
$U(1)$ transformations do. For $\lambda$ that is solely a function of time, it is possible to obtain a gauge invariant Lagrangian 
density by introducing a one-component gauge field $A_0$ and replacing the ordinary time derivatives of \eqref{4.1} with 
covariant time derivatives. However, as seen from the electric limit transformation formulas \eqref{1.4}, the introduction of a 
one-component gauge field $A_0$ is forbidden by Galilean invariance. The trouble with introducing a vector potential 
is the absence of a gradient operator
in the Lagrangian density to which it may couple, in turn a consequence of the UIR's of $\G(3:1)$. 
As we will see below, we may all the same introduce a vector potential in order to 
preserve Galilean invariance, but the absence of gradient operators in \eqref{4.1} rules out the existence 
of a current density that depends on the vector potential, and therewith also electrodynamics. The combined 
requirements of the unitary representations of 
the transformation group $\G(3:1)$ and $U(1)$-gauge invariance can only give rise to electrostatics, in contrast to what may 
have been anticipated from the general discussion of the Introduction based on the group $\G(3:1)$ itself, rather than 
its unitary representations. The situation is also in sharp contrast to the \emph{electrodynamics} in the  magnetic limit, 
based on the UIR's of the Galilei group 
$\G(1:3)$. 

\subsection{$\bs{U(1)}$ gauge transformations: electrostatics}\label{sec4.1} 
The structure of the Lagrangian density \eqref{4.1} demands the introduction of a gauge field $A_0$ in order to ensure local $U(1)$ 
gauge invariance. While not so mandated by the Lagrangian, we can nevertheless introduce a vector potential $\bs{A}$, and therewith 
a magnetic field $\bs{B}$, and work out the consequences. Our analysis will show that the structure of the dual Galilei 
group $\G(3:1)$ does not permit electrodynamics. 

We allow the matter field to undergo arbitrary gauge transformations 
\begin{eqnarray}
\psi(\bs{x},t) &\to& e^{-i\lambda(\bs{x},t)}\psi(\bs{x},t)\nonumber\\
\psi^*(\bs{x},t) &\to&e^{i\lambda(\bs{x},t)}\psi^*(\bs{x},t)\label{4.2}
\end{eqnarray}
and introduce scalar and vector (under rotations) potentials $A_0$ and $\bs{A}$ with the gauge transformation 
properties of \eqref{2.14}. In the position representation, we may write these operator equations simply as 
transformations of real valued functions of $(\bs{x},t)$: 
\begin{eqnarray}
A_0(\bs{x},t)\to \tilde{A}_0(\bs{x},t)&=&A_0(\bs{x},t)+\frac{1}{g}\frac{d\lambda(\bs{x},t)}{dt}\nonumber\\
\bs{A}(\bs{x},t)t\to \tilde{A}(\bs{x},t)&=&\bs{A}(\bs{x},t)+\frac{1}{g}\nabla\lambda(\bs{x},t)\label{4.3}
\end{eqnarray}
Then, with the usual definition \eqref{2.17} of the  gauge invariant $\bs{E}$ and $\bs{B}$ fields in terms of the derivatives of 
 $(A_0,\bs{A})$, we obtain from \eqref{4.1} the most general Lagrangian density that is invariant under  simultaneous 
 transformations \eqref{4.2} and \eqref{4.3}:
\begin{equation}
\L=\frac{i}{2}\psi^*\Bigl(\frac{d}{dt}+igA_0\Bigr)\psi-\frac{i}{2}\psi\Bigl(\frac{d}{dt}-igA_0\Bigr)\psi^*-E\psi^*\psi+f(\bs{E},\bs{B})\label{4.4}
\end{equation}
where $f$ arbitrary.
Note that, in contrast to  \eqref{2.17b}, here we do not have gauge covariant spatial derivatives, a peculiar feature 
directly related to the assumption that the matter field $\psi$ transforms irreducibly under the dual Galilei group $\G(3:1)$, rather than 
the Galilei group $\G(1:3)$.  

By way of Euler-Lagrange equations, we obtain from \eqref{4.4} the equations of motion: for the matter field $\psi$, 
\begin{equation}
i\Bigl(\frac{d}{dt}+igA_0\Bigr)\psi=E\psi,\label{4.5}
\end{equation}
for the gauge field $A_0$,
\begin{equation}
\nabla\cdot\nabla_{\bs{E}}f=-g\psi^*\psi,\label{4.6}
\end{equation}
and finally for $\bs{A}$,
\begin{equation}
\frac{d}{dt}\nabla_{\bs{E}}f+\nabla\times\nabla_{\bs{B}}f=0\label{4.7}
\end{equation}
Note that the matter field equation \eqref{4.5} differs from \eqref{2.14} in the absence of spatial gradients and field $\bs{A}$, 
while the gauge field equation \eqref{4.7} differs from the corresponding equation of \eqref{2.19} by the 
absence of a current density, which again is rooted in the absence of the gradient operators in the dynamical equation 
for the matter field.

The definitions of $\bs{E}$ and $\bs{B}$ fields \eqref{2.17} immediately imply the identities $\nabla\cdot\bs{B}=0$ and $\nabla\times\bs{E}+\frac{d\bs{B}}{dt}=0$. Now, if we set $f(\bs{E},\bs{B})=\frac{g^2}{2c}\left(\bs{E}^2-c^2\bs{B}^2\right)$, then together 
with \eqref{4.6} and \eqref{4.7}, we obtain Maxwell's equations:
\begin{eqnarray}
\nabla\cdot\bs{B}&=&0\nonumber\\
\nabla\times\bs{E}+\frac{d}{dt}\bs{B}&=&0\nonumber\\
\nabla\cdot\bs{E}&=&\frac{c}{g^2}\rho\nonumber\\
c^2\nabla\times\bs{B}-\frac{d}{dt}\bs{E}&=&0\label{4.8}
\end{eqnarray}
where 
\begin{equation}
\rho(\bs{x},t)=-g\psi^*(\bs{x},t)\psi(\bs{x},t)\label{4.9}
\end{equation}
These equations are to be supplemented by the matter field equation of motion \eqref{4.5}. 
It follows from \eqref{4.5}, or directly from the trivial time evolution of the wave function \eqref{3.25} that the charge density $\rho$ 
is time independent: 
\begin{equation}
\frac{d\rho}{dt}=0\label{4.10}
\end{equation}
This is of course the continuity equation in the absence of a current density. 

It is \eqref{4.10} that allows only for electrostatics here: it follows from \eqref{4.10} and Gauss's law that the $\bs{E}$ field must also be time independent. A static electric field implies static potentials and so in view of \eqref{2.17}, the $A_0$ and $\bs{A}$ must be such that all their time dependence is removable by a gauge transformation \eqref{4.3}. That is, $A_0$ and $\bs{A}$ must be of the form 
\begin{eqnarray}
A_0(\bs{x},t)&=&\tilde{A}_0(\bs{x})+\frac{d\lambda(\bs{x},t)}{dt}\nonumber\\
\bs{A}(\bs{x},t)&=&\tilde{\bs{A}}(\bs{x})+\nabla\lambda(\bs{x},t)\label{4.11}
\end{eqnarray}
for some function $\lambda$. 

It is clear from \eqref{4.11} that the magnetic field is also static. Therefore, we only have a curl free, divergence free, static magnetic field, and nothing 
interesting will happen magnetically.  With $\bs{B}$ static, from Amper\'e's law (or directly from \eqref{4.11}) we see
 that the $\bs{E}$ field is also curl free.

\subsection{Transformations of gauge field equations under $\bs{\G(3:1)}$}\label{sec4.2}
Definitions \eqref{2.17} and transformation formulas \eqref{1.4} show that we have constructed the $\bs{E}$ and $\bs{B}$ as components of a second rank tensor under the homogeneous Galilei group $\G(1:3)_{\rm hom}$.  In particular, in an arbitrary inertial frame defined by the dual Galilei group $\G(3:1)$, the transformed fields are 
\begin{eqnarray}
\bs{E}'(\bs{x},t)&=&R\bs{E}(\bs{x}')\nonumber\\
\bs{B}'(\bs{x},t)&=&R\bs{B}(\bs{x}')+\frac{\bs{\beta}}{c}\times R\bs{E}(\bs{x}')\label{4.12}
\end{eqnarray}
where $\bs{x}'$ is defined by \eqref{3.21}. 
Then, using \eqref{2.2.26} (Cf.~the duality between $d^\mu$ and $d_\mu$ in \eqref{1.17}), we obtain
\begin{eqnarray}
\nabla\cdot\bs{E}'(\bs{x},t)&=&R\nabla'\cdot\bs{E}(\bs{x}')=\rho(\bs{x}')\nonumber\\ 
c^2\nabla\times\bs{B}'(\bs{x},t)-\frac{d}{dt}\bs{E}'(\bs{x},t)&=&c^2R\nabla'\times\Bigl(R\bs{B}(\bs{x}')+\frac{\bs{\beta}}{c}\times R\bs{E}(\bs{x}')\Bigr)\nonumber\\
&&\ \ \ -\left(\frac{d}{dt'}-\bs{v}\cdot R\nabla'\right)R\bs{E}(\bs{x}')\nonumber\\
&=&\bs{v}\rho(\bs{x}')\label{4.13}
\end{eqnarray}
where the last equality follows from the third and fourth equalities of \eqref{4.8}. Now, defining 
\begin{eqnarray}
\rho'(\bs{x},t)&=&\rho(\bs{x}',t')\nonumber\\
\bs{j}'(\bs{x},t)&=&R\bs{j}(\bs{x}',t')+\bs{v}\rho(\bs{x}',t')\label{4.14}
\end{eqnarray}
we see that equations \eqref{4.13} have the same form as the last two equations of \eqref{4.8}. 

The transformation properties of the homogeneous equations can  be determined similarly. With the use of \eqref{3.21b}, 
\begin{eqnarray}
\nabla\cdot\bs{B}'(\bs{x}',t')&=&\left(R\nabla'+\frac{\bs{\beta}}{c}\frac{d}{dt'}\right)\cdot\left(R\bs{B}(\bs{x}')+\frac{\bs{\beta}}{c}\times R\bs{E}(\bs{x}')\right)\nonumber\\
&=&\nabla'\cdot\bs{B}(\bs{x}')+\frac{\bs{\beta}}{c}\cdot R\left(\frac{d}{dt'}\bs{B}(\bs{x}')-\nabla'\times\bs{E}(\bs{x}')\right)\nonumber\\
&=&0\nonumber\\
\nabla\times\bs{E}'(\bs{x},t)+\frac{d}{dt}\bs{B}'(\bs{x},t)&=&R\left(\nabla'\times\bs{E}(\bs{x}')+\frac{d}{dt'}\bs{B}(\bs{x}')\right)
+2\frac{d}{dt'}\left(\frac{\bs{\beta}}{c}\times R\bs{E}(\bs{x}')\right)\nonumber\\
&=&0\label{4.15}
\end{eqnarray}

Hence, Maxwell's equations have the same form in frames of reference connected by the dual group $\G(3:1)$. 
Note that the duality between the homogeneous and inhomogeneous equations  as well as the static nature of the 
fields are necessary to obtain \eqref{4.13} -- \eqref{4.15}. We see from \eqref{4.13} that there appears a current density that is simply proportional 
to the charge density.  Furthermore, it is also clear from \eqref{4.13} that it is possible to find a frame of reference in which the magnetic field 
vanishes, as expected of electrostatics.  

\section{Concluding remarks}\label{sec5}

As discussed in the Introduction, the covariance of a dynamical equation under a spacetime symmetry group is not uniquely determined by how spacetime variables transform under the given group, but depends in part on how various dynamical variables appearing in the equation are \emph{defined} to transform as tensors under the given group. As a simple example, note that Newton's second law can be stated as either an equation covariant under the Galilei group, $\bs{F}=m\bs{a}$, or as an equation covariant under the Lorentz group, $f^\mu=m\frac{d^2x^\mu}{d\tau^2}$. 
However, either equation acquires physical content only when supplemented with empirical `force laws', and a given set of force laws - expressed, say, as functions of spacetime variables - may have well defined transformation properties under one symmetry group but not the other. In other words, the statement that Newton's laws are consistent with Galilean relativity is true only for a class of force laws.

In the quantum mechanical setting, variables appearing in dynamical equations are often defined as functions of quantum state vectors (or fields). 
Such state vectors inhabit a Hilbert space which furnishes a unitary representation of the relevant spacetime symmetry group.  The central thesis advocated in this paper is that the transformation properties of dynamical variables inferred from unitary representations of spacetime symmetry groups have important implications for the covariance structure of the equations fulfilled by these dynamical variables. In addition, we have also shown that if the  homogeneous spacetime transformations are governed by a representation $D(g)$ of a symmetry group, then dynamical variables that couple to derivatives, such as gauge fields, must transform under the dual representation $C(g)=D^T\left(g^{-1}\right)$. Therefore, the analysis of the covariance of an equation that involves derivatives with respect to spacetime coordinates under a symmetry group $G$ must accommodate both $D$ and $C$ representations of $G$. 

Under this general theoretical construct, we have investigated two cases in detail: the magnetic limit and the electric limit of Maxwell's equations. 
If the charge and current densities are defined in terms of state vectors which transform under a unitary, projective, irreducible representation of the Galilei group, 
then a $U(1)$ gauge field that couples to derivative operators must be a vector field under the dual Galilei group $\G(3:1)$. Such a gauge field gives rise to 
electric and magnetic fields with meaningful transformation properties as a second rank tensor under 
$\G(3:1)$. This is the magnetic limit. Conversely, if the charge-current densities are defined in terms of state vectors which transform under a unitary 
irreducible representation of the dual Galilei group 
$\G(3:1)$, then a gauge field that couples to derivative operators must transform as a vector field under the Galilei group, \eqref{1.4}. This is the electric limit. 

The magnetic and electric limits were first introduce by Le Bellac and Levy-Leblond \cite{lebellac} as two distinct limits of the Lorentz 
transformation formula for the  tensor $F^{\mu\nu}$ 
for $\frac{v}{c}\to 0$.  We have shown here that the group theoretical content of these two limits is the existence of two In\"on\"u-Winger contractions of the Poincar\'e group, 
the usual one with respect to the time translation subgroup leading to the Galilei group $\G(1:3)$ and another with respect to the space translation subgroup leading to the dual Galilei 
group $\G(3:1)$. 

The main technical result we have reported in this paper is the construction of unitary, irreducible representations of the dual Galilei group $\G(3:1)$. We have shown that these representations are characterized by two Casimir operators which have interpretation as spin and energy. That energy is an invariant is a rather peculiar property related to the fact that time translations appear as a central subgroup of the dual Galilei group $\G(3:1)$. Thus, every state vector  being an eigenvector of the Hamiltonian with the same energy, a system described by a UIR of $\G(3:1)$ does not evolve in time. As an immediate consequence thereof, if charge-current densities are derived from a UIR of $\G(3:1)$, then only electrostatics is possible in the electric limit. 

That only a static solution is possible in the electric limit is in sharp contrast to the magnetic limit. It illuminates our central point that when dynamical variables are defined in terms of quantum state vectors, the covariance of an equation of motion and its dynamical content may be severely constrained by the representations of the symmetry group that defines the state vector space.

\section*{Acknowledgments} 
This research was supported by an award from Research Corporation. We thank Emily Moore, Victor Colussi and Anna McCoy for their support.

\end{document}